\documentclass[journal]{IEEEtran}

{}
{}
{}
\usepackage{amsmath,amsfonts,amssymb,amscd,amsthm,xspace}
\usepackage{lipsum}
\usepackage{url}
\newcommand{\PR}[1]{\ensuremath{\left[#1\right]}}
\newcommand{\PC}[1]{\ensuremath{\left(#1\right)}}
\newcommand{\chav}[1]{\ensuremath{\left\{#1\right\}}}
\newcommand{\PM}[1]{\ensuremath{\left|#1\right|}}
\newcommand{\norm}[1]{\left\lVert #1 \right\rVert}
\usepackage{blindtext}

\usepackage{graphicx}
\usepackage{textcomp}
\usepackage{xcolor}
\usepackage{algorithm}
\usepackage[noend]{algpseudocode}
\usepackage{lipsum}
\usepackage{comment}
\usepackage{makecell}
\usepackage{adjustbox}
\usepackage{multirow}
\usepackage{setspace}
\usepackage{soul}
\definecolor{r}{rgb}{0,0,0}
\definecolor{red}{rgb}{0,0,0}

\begin{document}

\title{Iterative Access Point Selection, MMSE Precoding and Power Allocation for Cell-Free Networks}

\author{Victoria M. T. Palhares, Rodrigo C. de Lamare, André R. Flores, Lukas T. N. Landau
\thanks{The authors are with the Center for Telecommunications Studies, Pontifical Catholic University of Rio de Janeiro, Rio de Janeiro, Brazil. Email: delamare@cetuc.puc-rio.br}}

\maketitle

\begin{abstract}
In this work, we propose iterative access point (AP) selection (APS), linear minimum mean-square error (MMSE) precoding and power allocation techniques for Cell-Free Massive multiple-input multiple-output (MIMO) systems. We consider the downlink channel with single-antenna users and multiple-antenna APs. We derive sum-rate expressions for the proposed iterative APS techniques followed by MMSE precoding and optimal, adaptive, and uniform power allocation schemes. Simulations show that the proposed approach outperforms existing conjugate beamforming (CB) and zero-forcing (ZF) schemes and that performance remains excellent with APS, in the presence of perfect and imperfect channel state information (CSI).
\end{abstract}

\section{Introduction}\label{sec1}

In the fifth generation (5G) of wireless communications, massive multiple-input multiple output (MIMO) systems \cite{Marzetta2010,mmimo,wence} have been used to improve the reliability, efficiency and throughput of wireless networks. In these systems, multiple antennas simultaneously serve  multiple users in the same time-frequency resource \cite{Marzetta2010}. In 5G, distributed antenna systems (DAS) are being considered in order to provide higher coverage probability, flexible resource management, higher power efficiency and larger capacities by the exploitation of smaller distances between base stations and users, and of spatial diversity \cite{Liu2014}. The combination of Massive MIMO and DAS systems has been exploited in the past, where it has been proven that distributed setups can provide higher rates than collocated antenna systems (CAS) \cite{Truong2013}. In \cite{Joung2014}, it is shown that large-scale DAS (L-DAS) systems are  energy-efficient and have simpler precoding and power control, when associated with antenna selection and user clustering methods.

As an evolution of Massive MIMO and DAS, in Cell-Free Massive MIMO, many randomly distributed access points (APs) are connected to a central processing unit (CPU) and serve simultaneously a much smaller number of users. At the CPU, precoding techniques and power allocation algorithms can be performed. The main goal is to use advanced backhaul in order to provide uniformly good service for all users \cite{Ngo2015}. Compared to a cellular system, cell-free concepts have been shown to increase energy efficiency and per-user throughput, in rural and urban scenarios \cite{Yang2018}. Moreover, they can have simple signal processing which facilitates the exploitation of phenomenons such as favorable propagation with channel hardening \cite{Marzetta2016}. \textcolor{r}{Nevertheless, channel hardening can be achieved in specific cell-free scenarios \cite{Chen2018}. If multiple-antenna APs are considered, channel hardening can be experienced at the cost of losses in macro-diversity. In a single-antenna APs scenario, increasing AP density does not lead to more channel hardening. An exception is made if a small \textcolor{red}{path loss} component is assumed, increasing the chances of this phenomenon with the increment in AP density.}

Many precoding and receive processing schemes have been previously
studied for cellular and cell-free networks
\cite{Landau,Spencer1,Stankovic,Sung,Zu_CL,Zu,wlbd,rsbd,rmmse,locsme,okspme,lrcc,mbthp,rmbthp,rsthp,baplnc,jpba,zfsec,rprec&pa}.
In the uplink, a matched filter (MF) has been exploited in
\cite{Ngo2017}, whereas in \cite{Bjornson2020,Nayebi2016}, a minimum
mean-square error (MMSE) and large-scale fading decoding (LSFD)
receivers have demonstrated to provided higher outage rate than the
former. \textcolor{r}{Moreover, in \cite{Chen2020}, high spectral
efficiency was achieved through an uplink framework combined with
local partial MMSE and maximum ratio (MR) combining.} On the other
hand, in the downlink, a conjugate beamforming (CB) precoder has
been studied for computational simple signal processing, at the cost
of lower performance \cite{Ngo2017,Nayebi2017,Ngo2017a}. With more
backhaul requirements, a zero-forcing (ZF) precoding design has been
extensively studied in order to improve system performance
\cite{Nayebi2017} and maximize the energy efficiency
\cite{Nguyen2017}. \textcolor{r}{In \cite{Interdonato2020}, two
distributed precoding schemes based on the ZF criterion have been
developed, which are called local partial ZF and local protective
partial ZF. In \cite{Bjornson2020a}, an MMSE-type precoder was
obtained through an uplink-downlink duality based on the MMSE
combiner from \cite{Bjornson2020}. The technique was combined with a
clustering method based on pilot assignment and \textcolor{red}{a}
heuristic solution for power allocation. Moreover, in a scenario
which considers  cell-free networks and non-orthogonal multiple
access (NOMA), precoders  based on MR transmission, full pilot ZF
precoding and modified regularized ZF precoding \cite{Rezaei2020}
have been evaluated.}

Since uniformly good service for all users is a key point of cell-free systems, many precoding and receive processing techniques have been combined with power control algorithms. Different criteria have been used in the literature to satisfy different aspects of this network concept. In \cite{Bashar2019}, power allocation is performed on the uplink to maximize the minimum user rate under per-user power constraints. In \cite{Ngo2017}, the MF and CB receiver and precoder, respectively, are combined with an optimization that maximizes the smallest of all user rate\textcolor{red}{s} under per-AP constraints. In \cite{Nayebi2017}, a similar procedure is performed, where CB and ZF precoders are combined with the maximization of the minimum signal-to-interference-plus-noise ratio (SINR) from all users also under per-AP power constraints. With a different criterion, ZF precoders were also combined with an energy efficiency maximization, under a per-AP power constraint and a per-user spectral efficiency constraint \cite{Ngo2017a} considering backhaul power consumption and imperfect channel state information (CSI) \cite{Nguyen2017}. \textcolor{r}{In \cite{Buzzi2020}, power control algorithms focused on maximizing the sum-rate or the minimum rate are developed and combined with a ZF precoder for cell-free systems.}

To decrease the backhaul power consumption in cell-free systems, some works have proposed AP selection (APS), where each user is served by a subset of APs. In \cite{Ngo2018}, two APS schemes have been proposed, one based on the  received power and the other on the largest large-scale-fading coefficients. \textcolor{r}{Similarly, a sequential approach to connect each user only to certain APs based on the channel gain and the channel quality has been proposed in \cite{Dao2020}.} Most works addressing Cell-Free Massive MIMO systems consider single-antenna APs. However, in \cite{Chen2018}, multiple-antenna APs have been analysed in order to improve channel hardening and to increase the likelihood that favorable propagation occurs. \textcolor{r}{In the same way, \cite{Mai2020} explores the downlink spectral efficiency in a scenario with multiple-antenna APs and users}. Additionally, in \cite{Ibrahim2017} it is shown that, in terms of costs, it is better to add more antennas to an AP than to install more APs.

In this work, we propose iterative APS, linear MMSE precoding and power allocation techniques \cite{aps&prec}. Specifically, APS, precoding and power allocation scheme\textcolor{r}{s are}  presented, where the precoder is calculated based on initial parameters, used in power allocation and recalculated based on power allocation coefficients. MMSE channel estimates are considered, similarly to previous works \cite{Nayebi2017}.
Furthermore, optimal, adaptive and uniform power allocation techniques are devised along optimal and suboptimal APS, and compared with existing CB and ZF precoding and power allocation techniques \cite{Ngo2017,Nayebi2017} in terms of sum-rate, minimum SINR and bit error rate (BER), taking into account perfect and imperfect CSI. Moreover, analytical expressions are derived to compute the achievable rates of the proposed approaches. Differently from \cite{Wang2015} which considers an asymptotic rate analysis in the number of antennas and users, we consider them to be finite.

In summary, the main contributions of this work are:
\vspace{-0.5em}

\begin{itemize}
    \item \textcolor{r}{Optimal APS scheme based on exhaustive search (ES) and a suboptimal solution based on the large-scale fading coefficients applied to the obtained channel parameters. In contrast with previous works, the optimal scenario has not been explored to date, and the proposed APS schemes have not been combined with the studied precoding and power allocation strategies.}
    \item \textcolor{r}{The complete derivation of an iterative linear MMSE precoder for Cell-Free Massive MIMO systems, which takes into account a power allocation matrix, unlike existing approaches \cite{Bjornson2020a,Joham2005}.}
    \item \textcolor{r}{Optimal and uniform power allocation techniques applied to the proposed scheme in order to maximize the minimum SINR, differently from \cite{Bjornson2020a} where the MMSE precoder is combined with a heuristic power allocation approach.}
    \item Adaptive power allocation based on the Stochastic Gradient (SG) with the mean-square error (MSE) criterion\textcolor{r}{, differently from existing power control solutions.}
    \item An analytical expression of the achievable rate\textcolor{r}{, specifically for the proposed MMSE precoder.}
    \item An analysis of the computational complexity of the proposed and existing schemes.
    \item \textcolor{r}{A simulation study of the proposed and existing techniques in terms of sum-rate} and minimum SINR. Unlike existing results from the literature, we illustrate the performance of the proposed system in terms of BER as well.
\end{itemize}

The rest of this paper is organized as follows. In Section II the Cell-Free Massive MIMO system model and CSI scenarios are detailed. In Section III, an iterative APS, combined with MMSE precoder and power allocation is presented. Sum-rate performance and the computational cost are evaluated in Section IV. In Section V, numerical results and discussions are presented, whereas in Section VI conclusions are drawn.

\textit{Notation}: Uppercase and bold symbols denote matrices and vectors respectively. The superscripts $\PC{}^*$,$\PC{}^T$, $\PC{}^H$ stand for complex conjugate, transpose and Hermitian  operations\textcolor{r}{,} respectively. The expectation, trace of a matrix, real part of the argument, Euclidean norm and Frobenius norm are denoted by $\mathbb{E}\PR{\cdot}$, $\text{tr}\PC{\cdot}$, $\text{Re}\PC{\cdot}$, $\norm{\cdot}_2$ and $\norm{ \cdot}_F$, respectively. The Hadamard product is expressed by $\odot$. The operator $\text{diag} \chav{\mathbf{A}}$ retains the main diagonal elements of $\mathbf{A}$ in a column vector. The $D \times D$ identity matrix is $\mathbf{I}_D$. The notation $x \sim \mathcal{N}(0,\sigma^2)$ refers to a Gaussian random variable (RV) $x$ with zero mean and variance $\sigma^2$ and $x \sim \mathcal{CN}(0,\sigma^2)$ denotes a circularly symmetric complex Gaussian RV $x$ with zero mean and variance $\sigma^2$.

\section{System Model}
\label{system_model}
\begin{figure}[!ht]
\centering
\includegraphics[width = 0.4\textwidth]{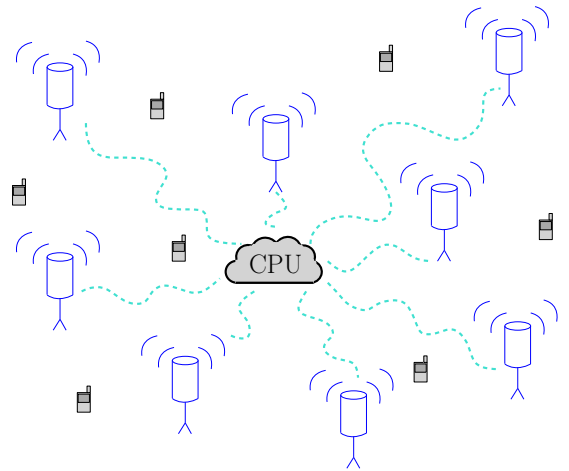}
\caption{Cell-Free Massive MIMO System.}\label{CF_Layout}
\end{figure}

The downlink of a Cell-Free Massive MIMO system is considered with $L$ randomly distributed APs equipped with $N$ antenna elements each and $K$ single-antenna users, where $LN = M$ is the total number of antenna elements and $M >> K$. In this system, all APs are connected to a CPU and serve simultaneously all users, as shown in Fig.~\ref{CF_Layout}. Each AP obtains CSI and sends them to the CPU, which performs APS, precoding and power allocation whose parameters are then fed back to the APs.

The channel coefficients between the $m$th antenna element and the $k$th user are defined as \cite{Ngo2017}
\begin{equation}
    g_{m,k} = \sqrt{\beta_{m,k}} h_{m,k},
    \label{def_g_mk}
\end{equation}
where $\beta_{m,k}$ is the large-scale fading (LS) coefficient (path loss and shadowing effects) and $h_{m,k} \sim \mathcal{CN}(0,1)$ is the small-scale fading coefficient, defined as independent and identically distributed (i.i.d) RVs that remain constant during a coherence interval and are independent over different coherence intervals. However, the LS coefficients change less frequently, being constant for several coherence intervals.   In this case, we assume that they change at least $40$ times slower than $h_{m,k}$ \cite{Ngo2017}. Assuming a carrier frequency of $1.9$ GHz and low user's mobility, the coherence interval is large and many orthogonal pilots can be used in channel estimation. Therefore, considering that the same pilot is assigned for users far away from each other, pilot contamination is assumed negligible \cite{Nayebi2017}.

Since each AP has $N$ antenna elements, we have for the $l$th AP
\vspace{-0.3em}
\begin{equation}
    \beta_{(l-1)\cdot N+1,k} = \beta_{(l-1)\cdot N+2,k} = \dots = \beta_{l\cdot N,k},~~ {\rm for}~ l=1,\dots,L,
\end{equation}
where we assume
that the links between the antenna elements of an AP and the users have the same distance and are affected by the same path loss and shadowing effects \cite{Chen2018,Ibrahim2017}.

In order to reduce interference between signals intended for different users, considering that each user is served by all APs, channel coefficients need to be taken into account when forming transmitted signals.
The system employs the time division duplex (TDD) protocol, where the channel is estimated through uplink training. First, all users send simultaneously and synchronously pilot sequences, $\boldsymbol{\Pi}_1,\dots,\boldsymbol{\Pi}_k,\dots,\boldsymbol{\Pi}_K \in \mathbb{C}^{\tau}$, to each antenna, where $\|\boldsymbol{\Pi}_k\|_{\textcolor{r}{2}}^2 = 1$. Then, APs perform MMSE channel estimation and obtain $\hat{g}_{m,k}$, the estimate of the channel coefficient $g_{m,k}$ between the $m$th antenna and the $k$th user. With $\hat{g}_{m,k}$, the data are transmitted to all users.

In the training step, the received signal sequence by the $m$th antenna is given by
\begin{equation}
    \mathbf{y}_{m} = \sqrt{\rho_r \tau} \sum_{k=1}^K g_{m,k} \boldsymbol{\Pi}_k + \mathbf{w}_m,
\end{equation}
where $\rho_r$ is the uplink power, $\tau$ is the length of pilot sequences and $\mathbf{w}_m \sim \mathcal{CN}(0,\mathbf{I}_{\tau})$ is the additive noise.

The MMSE estimate of $g_{m,k}$ is given by
\begin{equation}
    \hat{g}_{m,k} = \frac{\sqrt{\rho_r \tau} \beta_{m,k}}{1+\rho_r \tau \beta_{m,k}} \boldsymbol{\Pi}_k^H \mathbf{y}_m.
\end{equation}
Since
\begin{equation}
   \tilde{g}_{m,k} =  g_{m,k} -  \hat{g}_{m,k}
\end{equation}
where $\tilde{g}_{m,k}$ is the CSI error between the $m$th antenna element and the $k$th user, we define
\begin{equation}
    \hat{g}_{m,k} \sim \mathcal{CN}(0,\alpha_{m,k}) \text{,} \qquad \tilde{g}_{m,k} \sim \mathcal{CN}(0,\beta_{m,k} - \alpha_{m,k}),
\end{equation}
and
\begin{equation}
    \alpha_{m,k} = \frac{\rho_r \tau \beta^2_{m,k}}{1+\rho_r \tau \beta_{m,k}}.
\end{equation}
We notice that \textcolor{r}{$\alpha_{m,k}$ as well as $\hat{g}_{m,k}$}, are functions of $\beta_{m,k}$. Therefore, to evaluate perfect and different levels of imperfect CSI we consider $\alpha_{m,k}$ as an adjustable percentage of $\beta_{m,k}$ ($0 \leq$ \textcolor{r}{$n_{m,k}$} $\leq 1$). Thus, we have
\begin{equation}
\begin{split}
    \alpha_{m,k} &= \textcolor{r}{n_{m,k}} \beta_{m,k}\\
    \tilde{g}_{m,k} &= g_{m,k} - \hat{g}_{m,k}, \text{ and}\\
    \mathbb{E} \PR{\PM{\tilde{g}_{m,k}}^2} &= \PC{1-\textcolor{r}{n_{m,k}}} \beta_{m,k}.
    \end{split}
\end{equation}
We assume channel reciprocity, which means that the channel coefficients for the uplink and downlink are the same.
After channel estimation is completed, the data is transmitted to all users.
The signal received by the $k$th user is described by
\begin{equation}
    y_k = \sqrt{\rho_f} \ \mathbf{g}^T_k \mathbf{P} \mathbf{s} + w_k,
    \label{received_signal_k_1}
\end{equation}
where $\rho_f$ is the maximum transmit power of each antenna, $\mathbf{g}_k = \PR{g_{1,k},\dots,g_{M,k}}^T$ are the channel coefficients for user $k$, $\mathbf{P} \in \mathbb{C}^{M \times K}$ is a generic linear precoding matrix,  $\mathbf{s} = [s_1,\dots,s_K]^T$ is the zero mean symbol vector, with $\sigma_s^2 = \mathbb{E} (|s_k|^2)$, $s_k$ is the data symbol intended for user $k$ (uncorrelated between users), $w_k \sim \mathcal{CN}(0,\sigma_{w}^2)$ is the additive noise for user $k$ and $\sigma_{w}^2$ is the noise variance. Therefore, for all users combined, we have the model
\begin{equation}
    \mathbf{y} = \sqrt{\rho_f} \ \mathbf{G}^T \mathbf{P} \mathbf{s} + \mathbf{w},
    \label{received_signal_1}
\end{equation}
where $\mathbf{G} \in \mathbb{C}^{M \times K}$ is the channel matrix with elements [$\mathbf{G}$]$_{m,k} = g_{m,k}$ and $\mathbf{w}=\PR{w_1,\dots,w_K}^T$ is the noise vector.


\section {Proposed Iterative APS, MMSE Precoding and Power Allocation}
\label{mmse_ap_select_power_alloc}

In this section, we present the proposed iterative APS, linear MMSE precoding and power allocation techniques. In particular, the APS, precoding and power allocation tasks are integrated into a cell-free framework. Furthermore, APS algorithms based on an \textcolor{r}{ES} and LS coefficients, linear MMSE precoding and power allocation techniques based on max-min fairness and adaptive techniques are developed to improve the performance of existing approaches.

\subsection{APS Techniques}
\label{ap_selection}

We present two APS techniques for reducing power consumption, and consequently, increasing the energy efficiency in cell-free settings. First, we explain how to perform APS in the proposed and existing precoding techniques, taking into consideration the power allocation problems that are examined in this work. An \textcolor{r}{ES} APS (ES-APS) scheme and a suboptimal LS based APS (LS-APS) with low computational cost are devised \cite{Ngo2018}.

\subsubsection{Precoding with APS}
\label{precod_ap_selection}

To perform APS with precoding, different approaches need to be used in the optimization problems. First, an auxiliary matrix $\mathbf{Q} \in \mathbb{N}^{M \times K}$, where [$\mathbf{Q}$]$_{m,k} = q_{m,k}\in \chav{0,1}$ is considered in order to define which links between APs and users are taken into account (as 1) and the ones that are discarded (as 0). Since we perform APS rather than  select antenna elements, all antennas from the $l$th AP should be considered or discarded. Therefore, the same procedure carried out for the LS coefficients ($\beta_{m,k}$) is done for $q_{m,k}$ where, $q_{(l-1)\cdot N+1,k} = q_{(l-1)\cdot N+2,k} = \dots = q_{l\cdot N,k}$, for $l=1,\dots,L$.

Before calculating the MMSE precoder, each coefficient $q_{m,k}$ will be multiplied by $\alpha_{m,k}$, $\beta_{m,k}$, $\hat{g}_{m,k}$ and $\tilde{g}_{m,k}$ so that $\alpha^{'}_{m,k} = q_{m,k} \cdot \alpha_{m,k}$, $\beta^{'}_{m,k} = q_{m,k} \cdot \beta_{m,k}$, $\hat{g}^{'}_{m,k} = q_{m,k} \cdot \hat{g}_{m,k}$ and $\tilde{g}^{'}_{m,k} = q_{m,k} \cdot \tilde{g}_{m,k}$. Then, precoding and power allocation algorithms are performed with these parameters. Similarly, we apply the same technique to the ZF precoder from \cite{Nayebi2017}. On the other hand, to perform APS in the CB precoder from \cite{Ngo2017,Nayebi2017}, the max-min fairness power allocation algorithm turns into a mixed continuous/discrete optimization problem \cite{Clarke2012}.

 \subsubsection{Exhaustive Search Selection (ES-APS)}
 \label{es_aps}
In this scheme, all possible sets considering a total of $L$ APs, $S$ selected APs and $K$ users, will be tested so that the combination that maximizes the minimum SINR is chosen. Each possible vector will be a column of matrix $\mathbf{V}$, formed by $S$ 1s and $(L-S)$ 0s. By choosing a set of $K$ columns of matrix $\mathbf{V}$, performing all combinations between them, $\binom{L}{S}^K$, and by replicating every row $N$ times to represent the selection of all $N$ antenna elements of a certain AP, we form all possibilities for $\mathbf{Q}$.

The proposed optimization problem that performs APS using ES is given by
\begin{subequations}
\begin{align}
&\max_{\mathbf{Q}} \min_{k}\ \text{SINR}_{k}\\
&\text{s.t.} \sum_{j=1}^{M} q_{j,k} = SN, \ k=1,\dots,K,
\end{align}
\end{subequations}
\textcolor{r}{where the SINR for user $k$ reads as}
\begin{equation}
\text{SINR}_{k} = \frac{\mathbb{E} \PR{|A_1|^2}}{\sigma_w^2 + \sum_{i=1,i \neq k}^{K} \mathbb{E} \PR{|A_{2,i}|^2}+\mathbb{E} \PR{|A_3|^2}},
\label{eq:SINR_k}
\end{equation}
\textcolor{r}{which is different to the expression in \cite{Nayebi2017} that relies on ZF precoding explicitly. In equation \eqref{eq:SINR_k}
\begin{equation}
    A_1 = \sqrt{\rho_f} \hat{\mathbf{g}}^{T}_k \mathbf{p}_{k} s_k,
\end{equation}
describes the desired signal, $\sigma_w^2$ is the noise variance,
\begin{equation}
    A_{2,i} = \sqrt{\rho_f} \hat{\mathbf{g}}^{T}_k \mathbf{p}_{i} s_i,
\end{equation}
is the interference caused by user $i \ \textrm{for} \ i \neq k, i = 1,\dots,K$,
\begin{equation}
    A_3 = \sqrt{\rho_f} \tilde{\mathbf{g}}^{T}_k \mathbf{P} \mathbf{s}
\end{equation}
is the CSI error, all based on \eqref{received_signal_k_1}, and $q_{j,k}$ is the $j$th element of the $k$th column of $\mathbf{Q}$. }\textcolor{r}{
Note that for the ES-APS approach, the whole procedure is done, including the precoding and power allocation steps, for each possibility. After calculation of the minimum SINRs for all arrangements, we choose the setup that yields the maximum minimum SINR.
}

\subsubsection{Large-Scale-Fading-Based Selection with Fixed Number of APs (LS-APS)}
 \label{ls_aps}

As an alternative to ES-APS, we devise a LS based selection method, where user $k$ will only be associated with $S \leq L$ APs corresponding to the $SN$ largest LS coefficients for user $k$, inspired by the algorithm in \cite{Ngo2018}. Unlike the approach in \cite{Ngo2018} that chooses the number of selected APs based on their contribution to the sum of the LS coefficients, LS-APS sets a fixed $S$ and then performs APS. This allows a fair comparison with ES-APS.

\begin{algorithm}[ht!]
\caption{LS-APS }
\begin{algorithmic}[1]
\State Estimate $\boldsymbol{\beta}_k = \PR{\beta_{1,k},\dots,\beta_{M,k}}^T$ for user $k$ and sort the elements in descending order.
\State Choose the largest $SN$ elements of $\boldsymbol{\beta}_k$.
\State In the auxiliary vector $\mathbf{q}_k$, assign 1s to the corresponding $SN$ largest elements of $\boldsymbol{\beta}_k$ and 0s to the remaining ones.
\State Let $\mathbf{q}_k$, $k=1,\dots,K$ be the columns of the matrix $\mathbf{Q}$.
\State Let $\alpha^{'}_{m,k} = q_{m,k} \cdot\alpha_{m,k}$, $\beta^{'}_{m,k} = q_{m,k} \cdot \beta_{m,k}$, $\hat{g}^{'}_{m,k} = q_{m,k} \cdot \hat{g}_{m,k}$ and $\tilde{g}^{'}_{m,k} = q_{m,k} \cdot \tilde{g}_{m,k}$, where $q_{m,k}$ is the $m$th element of vector $\mathbf{q}_k$.
\State Perform the precoding and power allocation with the new parameters.
\end{algorithmic}
\label{algorithm_1}
\end{algorithm}

To perform APS with LS-APS, as detailed in Algorithm \ref{algorithm_1}, we first need to estimate $\boldsymbol{\beta}_k = \PR{\beta_{1,k},\dots,\beta_{M,k}}^T$ for user $k$ and sort the elements in descending order. Then, we assign in an auxiliary vector $\mathbf{q}_k$ 1s corresponding to the $SN$ largest elements of $\boldsymbol{\beta}_k$ and 0s to the remaining entries. By grouping $\mathbf{q}_k$ as the columns of $\mathbf{Q}$, we obtain the matrix $\mathbf{Q}$. This completes the APS stage.

\subsection{Precoder Design}
\label{precoder_design}
In this subsection, we detail the derivation of the proposed linear MMSE precoder. Unlike existing approaches \cite{Nayebi2017,Joham2005}, in the proposed precoding technique, we consider power allocation in the derivation and take into account the CSI matrix after APS, $\hat{\mathbf{G}}^{'}$, instead of the actual channel matrix $\mathbf{G}$, since the APs have imperfect CSI. Furthermore, we consider an iterative MMSE precoder with power allocation in order to maximize the minimum SINR. If the conventional MMSE precoder in \cite{Joham2005} was applied to this type of system, the performance would be degraded due to the lack of appropriate power allocation. In the precoder design we also take into account a normalization factor $f^{-1}$ at the receivers, which can be interpreted as an automatic gain control \cite{Joham2005}.

By modifying the expression in \eqref{received_signal_k_1}, taking into account both the precoding and the power allocation matrices, the signal received by the $k$th user is given by
\begin{equation}
    y_k = \sqrt{\rho_f} \ \mathbf{g}^T_k \mathbf{P} \ \mathbf{N} \ \mathbf{s} + w_k,
    \label{received_signal_k}
\end{equation}
where $\mathbf{N} \in \mathbb{R}_{+}^{K \times K}$ is the power allocation diagonal matrix with $\sqrt{\eta_1},\dots,\sqrt{\eta_K}$ on its diagonal and $\eta_k$ is the power coefficient of user $k$. Therefore, for all users combined, we have
\begin{equation}
    \mathbf{y} = \sqrt{\rho_f} \ \mathbf{G}^T \mathbf{P} \ \mathbf{N} \mathbf{s} + \mathbf{w},
    \label{received_signal}
\end{equation}
To obtain the proposed MMSE precoder, the following optimization is solved \cite{Joham2005}

\begin{subequations}\label{opt_problem}
    \begin{align}
            &\chav{\mathbf{P}_{\text{MMSE}},\mathbf{N},f_{\text{MMSE}}} = \text{argmin}_{\chav{\mathbf{P},\mathbf{N},f}} \mathbb{E} \PR{\norm{\mathbf{s}-f^{-1}\mathbf{y}}_2^2} \label{a}\\
            & \quad \text{s.t.: } \mathbb{E} \PR{\norm{\mathbf{x}}_2^2} = E_{tr},
    \end{align}
\end{subequations}

where the transmitted signal is given by
\begin{equation}
    \mathbf{x} = \sqrt{\rho_f} \ \mathbf{P} \ \mathbf{N} \ \mathbf{s}.
\end{equation}
The average transmit power is described by
\begin{equation}\label{eq_total_power}
    \mathbb{E} \PR{\norm{\mathbf{x}}_2^2} = \rho_f \text{tr}\PC{\mathbf{P} \mathbf{N} \mathbf{C}_s \mathbf{N}^H \mathbf{P}^H} = E_{tr},
\end{equation}
where $\mathbf{C}_s = \mathbb{E} \PR{\mathbf{s}\mathbf{s}^H}$ is the symbol covariance matrix.

By constructing the Lagrangian function with the Lagrange multiplier $\lambda$, setting its derivatives with respect to the precoder and the normalization factor to zero, and considering a power allocation matrix $\mathbf{N}$, we can compute the precoder ${\mathbf P}$ and the normalization $f$, as shown below:
\begin{equation}
\begin{split}
\mathcal{L} \PC{\mathbf{P},\mathbf{N},f, \lambda}  &= \ \mathbb{E} \PR{\norm{\mathbf{s}-f^{-1}\mathbf{y}}_2^2} + \lambda \left( \rho_f \text{tr} \PC{\mathbf{P} \mathbf{N} \mathbf{C}_s \mathbf{N}^H \mathbf{P}^H}\right.\\
&\left.- E_{tr}\right) = \text{tr} \PC{\mathbf{C}_s} - f^{-1} \sqrt{\rho_f} \text{tr} \PC{\hat{\mathbf{G}}^{'T} \mathbf{P} \mathbf{N} \mathbf{C}_s}\\
&-  f^{-1} \sqrt{\rho_f} \text{tr} \PC{\hat{\mathbf{G}}^{'*}\mathbf{C}_s \mathbf{N}^H \mathbf{P}^H} + f^{-2} \text{tr} \PC{\mathbf{C}_w}\\
&+ f^{-2} \rho_f \text{tr} \PC{\hat{\mathbf{G}}^{'*} \hat{\mathbf{G}}^{'T} \mathbf{P} \mathbf{N} \mathbf{C}_s \mathbf{N}^H \mathbf{P}^H}\\
& + \lambda \PC{\rho_f \text{tr} \PC{\mathbf{P} \mathbf{N} \mathbf{C}_s \mathbf{N}^H \mathbf{P}^H} - E_{tr}},\\
\end{split}
\end{equation}
where $\mathbf{C}_w = \mathbb{E} \PR{\mathbf{w}\mathbf{w}^H}$ is the noise covariance matrix.

Using Wirtinger's calculus and the result of the partial derivative $\partial \text{tr} \PC{\mathbf{B} \mathbf{X}^H}/ \partial \mathbf{X}^{*} = \mathbf{B}$, we have
\begin{equation} \label{deriv_p}
\begin{split}
    \frac{\partial \mathcal{L} \PC{\mathbf{P},\mathbf{N},f, \lambda}}{\partial \mathbf{P}^{*}} &= f^{-2} \rho_f \hat{\mathbf{G}}^{'*} \hat{\mathbf{G}}^{'T} \mathbf{P} \mathbf{N} \mathbf{C}_s \mathbf{N}^H\\
    &+ \lambda \rho_f \left(\mathbf{P} \mathbf{N}\mathbf{C}_s \mathbf{N}^H\right)-f^{-1} \sqrt{\rho_f} \hat{\mathbf{G}}^{'*} \mathbf{C}_s \mathbf{N}^H  = 0
\end{split}
\end{equation}
and
\begin{equation} \label{deriv_f}
\begin{split}
    \frac{\partial \mathcal{L} \PC{\mathbf{P},\mathbf{N},f, \lambda}}{\partial f} & = 2 f^{-3} \text{tr} \left(-\left(\rho_f  \hat{\mathbf{G}}^{'T} \mathbf{P} \mathbf{N} \mathbf{C}_s \mathbf{N}^H \mathbf{P}^H \hat{\mathbf{G}}^{'*} \right.\right. \\
    &+ \left. \left.\mathbf{C}_w \right)+ f \sqrt{\rho_f} \ \text{Re} \PC{\hat{\mathbf{G}}^{'T} \mathbf{P} \mathbf{N} \mathbf{C}_s}\right) = 0.\\
\end{split}
\end{equation}
Solving for \eqref{deriv_p}, we obtain
\begin{equation}\label{eq_p_tilde}
    \mathbf{P}= \frac{f}{\sqrt{\rho_f}} \underbrace{\PC{\hat{\mathbf{G}}^{'*} \hat{\mathbf{G}}^{'T} + \lambda f^2  \mathbf{I}_M}^{-1} \hat{\mathbf{G}}^{'*}}_{\tilde{\mathbf{P}}} \mathbf{N}^{-1} = \frac{f}{\sqrt{\rho_f}} \  \tilde{\mathbf{P}} \mathbf{N}^{-1}.
\end{equation}
By using the expression in \PC{\ref{deriv_f}}, we arrive at
\begin{equation}
\resizebox{1\hsize}{!}{
    \text{tr} \PC{f^2 \hat{\mathbf{G}}^{'T} \tilde{\mathbf{P}} \mathbf{C}_s \tilde{\mathbf{P}}^H \hat{\mathbf{G}}^{'*} + \mathbf{C}_w - f^2 \text{Re} \PC{\hat{\mathbf{G}}^{'T} \tilde{\mathbf{P}} \mathbf{C}_s}} = 0}.
    \label{epsilon_exp}
\end{equation}
Considering $\epsilon = \lambda f^2$ and using the relation presented in \cite{Joham2005}, described by
\begin{equation}
\resizebox{0.95\hsize}{!}{
    \text{tr} \PC{\text{Re} \PC{\hat{\mathbf{G}}^{'T} \tilde{\mathbf{P}} \mathbf{C}_s}} = \text{tr} \PC{\PC{\hat{\mathbf{G}}^{'*} \hat{\mathbf{G}}^{'T} + \epsilon \ \mathbf{I}_M} \tilde{\mathbf{P}} \mathbf{C}_s \tilde{\mathbf{P}}^H}},
    \label{relation_6}
\end{equation}
we obtain

\begin{equation}
\text{tr} \PC{\mathbf{C}_w} - \epsilon \ f^2 \text{tr} \PC{\tilde{\mathbf{P}} \mathbf{C}_s \tilde{\mathbf{P}}^H} = 0 \\
\end{equation}
\begin{equation}
\text{tr} \PC{\mathbf{C}_w} - \epsilon E_{tr} = 0\\
\end{equation}
\begin{equation}
\epsilon = \frac{\text{tr} \PC{\mathbf{C}_w}}{E_{tr}} = \lambda f^2.
\end{equation}
\textcolor{r}{As shown above, $\lambda$ is not directly required for finding the optimal solution. With this, the auxiliary precoding matrix reads as
\begin{equation}
    \tilde{\mathbf{P}}=\PC{\hat{\mathbf{G}}^{'*} \hat{\mathbf{G}}^{'T} + \frac{\text{tr} \PC{\mathbf{C}_w}}{E_{tr}}  \mathbf{I}_M}^{-1} \hat{\mathbf{G}}^{'*}.
\end{equation}}
\textcolor{r}{
By inserting \eqref{eq_p_tilde} in \eqref{eq_total_power}
\begin{equation}
\begin{split}
E_{tr} &= \rho_f \text{tr}\PC{\mathbf{P} \mathbf{N} \mathbf{C}_s \mathbf{N}^H \mathbf{P}^H}\\
&= \rho_f \text{tr}\PC{\frac{f}{\sqrt{\rho_f}} \  \tilde{\mathbf{P}} \mathbf{N}^{-1} \mathbf{N} \mathbf{C}_s \mathbf{N}^H \PC{\mathbf{N}^{-1}}^H \tilde{\mathbf{P}}^H \frac{f}{\sqrt{\rho_f}}}\\
&= f^2\text{tr}\PC{\tilde{\mathbf{P}}\mathbf{C}_s\tilde{\mathbf{P}}^H},\\
\end{split}
\end{equation}
we find that
\begin{equation}
\begin{split}
f_{\text{MMSE}}&= \sqrt{\frac{E_{tr}}{\text{tr}\PC{\tilde{\mathbf{P}}  \mathbf{C}_s \tilde{\mathbf{P}}^H}}}.
\end{split}\label{exp_f_mmse}
\end{equation}}

Therefore, the proposed MMSE precoder that takes into account power allocation for cell-free systems is given by
\begin{equation}
    \mathbf{P}_{\text{MMSE}} = \frac{f_{\text{MMSE}}}{\sqrt{\rho_f}} \PC{\hat{\mathbf{G}}^{'*} \hat{\mathbf{G}}^{'T} + \frac{\text{tr} \PC{\mathbf{C}_w}}{E_{tr}}  \mathbf{I}_M}^{-1} \hat{\mathbf{G}}^{'*}\mathbf{N}^{-1},
     \label{mmse_precoder_opt}
\end{equation}
where $[\hat{\mathbf{G}}^{'}]_{m,k} = \hat{g}^{'}_{m,k}$ is the CSI matrix after APS and we consider $\text{tr} \PC{\mathbf{C}_w} = K \sigma_{w}^2$. We initialize the precoder considering a power allocation matrix $\mathbf{N} = \mathbf{I}_K$. After the MMSE precoder is obtained, we perform power allocation. With the new power allocation matrix $\mathbf{N}$, we substitute it in $\mathbf{P}_{\text{MMSE}}$. The last step of the iteration is to recalculate the matrix $\mathbf{N}$ by employing $\mathbf{P}_{\text{MMSE}}$ to perform power allocation. The final power allocation matrix, denoted as $\mathbf{N}_{\text{MMSE}}$, ensures that the power constraint is satisfied. Note that the $\mathbf{N}$ present in the $\mathbf{P}_{\text{MMSE}}$ expression is different from the final $\mathbf{N}_{\text{MMSE}}$. Therefore, $\mathbf{N}$ and $\mathbf{N}_{\text{MMSE}}$ will not cancel each other.

\subsection{Power Allocation}
 \label{power_alloc}

We present Optimal Power Allocation (OPA), Adaptive Power Allocation (APA) and Uniform Power Allocation (UPA) techniques to obtain $\mathbf{N}$, a diagonal matrix with $\sqrt{\eta_1},\dots,\sqrt{\eta_K}$ on its main diagonal, which is used to recompute the precoder $\mathbf{P}_{\text{MMSE}}$ and find the matrix $\mathbf{N}_{\text{MMSE}}$.

\subsubsection{Optimal Power Allocation (OPA)}
\label{opt_power_alloc}

The max-min fairness power allocation problem with antenna constraint is given by
\begin{subequations}
\begin{align}
&\max_{\boldsymbol{\eta}} \min_{k} \text{SINR}_{k} \PC{\boldsymbol{\eta}} \label{sinr_mmse_complete}\\
&\text{s.t.} \sum_{i=1}^K \eta_{i} \delta_{m,i} \leq 1, m=1,\dots,M,
\end{align}
\end{subequations}
where
\begin{equation}
    \boldsymbol{\delta}_m = \mathrm{diag} \chav{ \mathbb{E} \PR{\mathbf{p}_{m}^T \mathbf{p}_{m}^{*}}},
\end{equation}
$\delta_{m,i}$ is the $i$th element of vector $\boldsymbol{\delta}_m$ and $\mathbf{p}_{m} = \PR{p_{m,1},\dots,p_{m,K}}$ is the $m$th row of the precoder $\mathbf{P}_{\text{MMSE}}$.

The optimization problem in an epigraph form employs the bisection method at each step and is described by
\begin{subequations}\label{epigraph_mmse}
\begin{align}
& \text{find } \boldsymbol{\eta} \\
&\text{s.t.} \ \text{SINR}_{\textcolor{r}{k}} \PC{\boldsymbol{\eta}} \geq t, \ k=1,\dots,K, \label{sinr_epigraph_mmse}\\
& \qquad \sum_{i=1}^K \eta_{i} \delta_{m,i} \leq 1, \ m=1,\dots,M,
\end{align}
\end{subequations}
where $t = \frac{t_b + t_e}{2}$ is the midpoint of a chosen interval $(t_b,t_e)$ that contains the optimal value $t^{*}$, as in \cite{Boyd2019}. The value $T_{\text{OPA}}$ is the number of iterations for the bisection method.

\subsubsection{Adaptive Power Allocation (APA)}

We propose an adaptive SG learning algorithm, which takes into account the gradient of the error in \eqref{opt_problem} to  perform \textcolor{r}{APA} and has a per-antenna power constraint. Our main objective here is to propose alternatives to the OPA and UPA algorithms. The proposed APA technique aims to adjust the power coefficients $\eta_k$ so that they minimize the effect of the interference at the received signal vector $\mathbf{y}$. First, an unconstrained optimization is performed in accordance with \eqref{exp_cost_apa} to minimize the interference. Then, a per-antenna power constraint is applied in order to fulfill the cell-free network transmit power requirements.

To start with the optimization we will first calculate the cost function:
\begin{equation}
\begin{split}\label{exp_cost_apa}
\mathcal{C}( \textcolor{r}{\mathbf{N}}) & =  \mathbb{E} \PR{\norm{\mathbf{s}-f^{-1}\mathbf{y}}_2^2}= \text{tr} \PC{\mathbf{C}_s}+ f^{-2} \text{tr} \PC{\mathbf{C}_w}\\
&- f^{-1} \sqrt{\rho_f} \text{tr} \left(\hat{\mathbf{G}}^{'T} \textcolor{r}{\mathbf{P}_{\text{MMSE}}} \textcolor{r}{\mathbf{N}}\mathbf{C}_s\right)
\\
&-  f^{-1} \sqrt{\rho_f}\text{tr} \PC{\textcolor{r}{\mathbf{P}^H_{\text{MMSE}}} \hat{\mathbf{G}}^{'*} \mathbf{C}_s  \textcolor{r}{\mathbf{N}}^H }\\
&+ f^{-2} \rho_f \text{tr} \PC{\textcolor{r}{\mathbf{P}^H_{\text{MMSE}}} \hat{\mathbf{G}}^{'*} \hat{\mathbf{G}}^{'T} \textcolor{r}{\mathbf{P}_{\text{MMSE}}} \textcolor{r}{\mathbf{N}}\mathbf{C}_s  \textcolor{r}{\mathbf{N}}^H}.
\end{split}
\end{equation}
The instantaneous gradient with respect to $\mathbf{N}^{*}$ is
\begin{equation}
\begin{split}
\hat{\underset{\mathbf{N}^{*}}{\nabla}} \mathcal{C}\PC{\textcolor{r}{\mathbf{N}}} & = -  f^{-1} \sqrt{\rho_f} \textcolor{r}{\mathbf{P}^H_{\text{MMSE}}} \hat{\mathbf{G}}^{'*} \mathbf{C}_s\\
&+ f^{-2} \rho_f \textcolor{r}{\mathbf{P}^H_{\text{MMSE}}} \hat{\mathbf{G}}^{'*} \hat{\mathbf{G}}^{'T}\textcolor{r}{\mathbf{P}_{\text{MMSE}}} \textcolor{r}{\mathbf{N}} \mathbf{C}_s.
\end{split}
\end{equation}
Therefore, the matrix $\textcolor{r}{\mathbf{N}}$ is updated by
\begin{equation}
\begin{split}
\textcolor{r}{\mathbf{N}}[i+1] &= \textcolor{r}{\text{Re}} \left(\textcolor{r}{\mathbf{N}}[i] - \mu \left(-f^{-1} \sqrt{\rho_f} \textcolor{r}{\mathbf{P}^H_{\text{MMSE}}} \hat{\mathbf{G}}^{'*} \mathbf{C}_s\right.\right.\\
&\left. \left.\quad + f^{-2} \rho_f \textcolor{r}{\mathbf{P}^H_{\text{MMSE}}} \hat{\mathbf{G}}^{'*} \hat{\mathbf{G}}^{'T} \textcolor{r}{\mathbf{P}_{\text{MMSE}}} \textcolor{r}{\mathbf{N}}[i] \mathbf{C}_s\right)\right).
\end{split}
\end{equation}

\textcolor{r}{As previously stated, the updated $\mathbf{N}$ matrix will be calculated based on the initial MMSE precoder, $\mathbf{P}_{\text{MMSE}}$, which considered $\mathbf{N} = \mathbf{I}_K$. With this new $\mathbf{N}$, we substitute it in $\mathbf{P}_{\text{MMSE}}$. Finally, with the new $\mathbf{P}_{\text{MMSE}}$ matrix, we calculate the final $\mathbf{N}$, called $\mathbf{N}_{\text{MMSE}}$. It is important to emphasize that in every calculation of $\mathbf{N}$, the per-antenna power constraint is reinforced.}
\newpage
In Algorithm \ref{algorithm_2}, the proposed adaptive SG learning strategy that performs APA is explained in further detail.

\begin{algorithm}[h!]
\caption{SG Adaptive Power Allocation Algorithm}
\begin{algorithmic}[1]
\State \textbf{Parameters:} $\mu$ (step size) and $T_{\text{APA}}$ (number of iterations).
\State \textbf{Initialization:} $\textcolor{r}{\eta_{k}}[0] = 10^{-3}, k=1,\dots,K.$
\State \textbf{For} i= 0:$T_{\text{APA}}$
\State \quad Set $\textcolor{r}{\mathbf{N}}[i]$ as a diagonal matrix with $\sqrt{\textcolor{r}{\boldsymbol{\eta}}[i]}$ on its diagonal.
\State \quad Define $\mathcal{C}\PC{\textcolor{r}{\mathbf{N}}} = \PR{\norm{\mathbf{s}-f^{-1}\mathbf{y}}_2^2}$.
\State \quad Compute  $\hat{\underset{\mathbf{N}^{*}}{\nabla}} \mathcal{C}\PC{\textcolor{r}{\mathbf{N}}}$.
\State \quad \textcolor{r}{Calculate} $\textcolor{r}{\mathbf{N}}[i+1] = \textcolor{r}{\mathbf{N}}[i] - \mu \hat{\underset{\mathbf{N}^{*}}{\nabla}} \mathcal{C}\PC{\textcolor{r}{\mathbf{N}}}$.
\State \quad Obtain $\textcolor{r}{\mathbf{N}}[i+1] = \text{Re}\PC{\textcolor{r}{\mathbf{N}}[i+1] \odot \mathbf{I}_K}$.
\State \quad  \textcolor{r}{Compute $\boldsymbol{\eta}_k[i+1]=\vert \mathbf{N}_{k,k}[i+1]  \vert^2$, $k=1,\dots,K$}.
\State \quad \textcolor{r}{Scale according to} the per-antenna \textcolor{r}{power constraint} $\textcolor{r}{\boldsymbol{\delta}}_{m} \cdot \boldsymbol{\eta}[i+1]  \leq 1, m=1,\dots,M$ to adjust $\textcolor{r}{\mathbf{N}}[i+1]$
\State \textbf{end}
\State \textcolor{r}{Obtain $\mathbf{N} = \mathbf{N}[i+1]$}.
\end{algorithmic}
\label{algorithm_2}
\end{algorithm}

\begin{figure}[!ht]
\centering
\includegraphics[width=0.84\columnwidth]{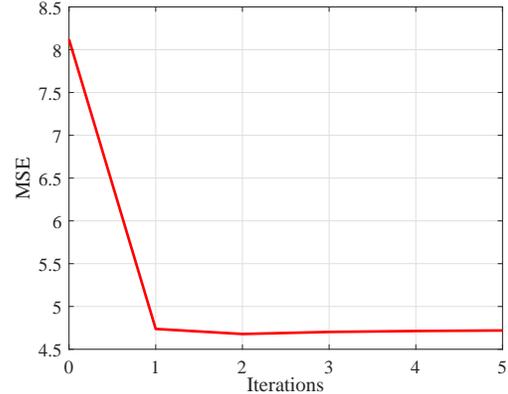}
\caption{Learning Algorithm - $L = 24$, $N = 4$, $S = 12$, $K = 8$\textcolor{r}{,} $n = 1$, SNR $= 25$ dB, $\mu = 0.25$ and $E_{tr} = M\rho_f$.} \label{Learning_APA}
\end{figure}

The proposed SG power allocation algorithm for APA has a  complexity of $\mathcal{O}(T_{\text{APA}}MK^2)$ and converges within only a few iterations. In Fig.~\ref{Learning_APA} we illustrate the performance of the proposed APA algorithm using a step size of $\mu = 0.25$ that converges within $T_{\text{APA}} = 5$ iterations.
\vspace{-1em}

\subsubsection{Uniform Power Allocation (UPA)}
\label{unif_power_alloc}
As an alternative to APA, an UPA scheme is proposed, based on the one in \cite{Nayebi2017} where a per-antenna power constraint is taken into account. Consider a case where we wish to obtain equal $\eta_k$ and \textcolor{r}{that} a certain antenna element $m$ transmits with full power. Then, with $\eta_k$ as its minimum possible value, we have
\begin{align}
\eta_k &= 1 / \PC{\max_{m} \sum_{i=1}^K \delta_{m,i}}, \ k=1,\dots,K, \label{upa_exp}
\end{align}
where $\delta_{m,i}$ is the $i$th element of vector $\boldsymbol{\delta}_m$. Although \eqref{upa_exp} is a suboptimal solution, it has low complexity, is simple and has the potential to show the benefits of the MMSE precoder.

\section{Analysis}
\label{sum_rate_analysis}
In this section, we present a sum-rate analysis of the proposed techniques along with the computational complexity of the proposed and existing algorithms.
\subsection{Sum-Rate}

First, we will expand expressions \eqref{received_signal_k} and \eqref{mmse_precoder_opt} to obtain the received signal by user $k$:
\begin{equation}
\begin{split}
y_k &= \sqrt{\rho_f} \ \mathbf{g}^{T}_k \mathbf{P}_{\text{MMSE}} \ \mathbf{N}_{\text{MMSE}} \ \mathbf{s} + w_k \\
&= \sqrt{\rho_f} \ \PC{\hat{\mathbf{g}}^{'}_k + \tilde{\mathbf{g}}^{'}_k}^T \frac{f_{\text{MMSE}}}{\sqrt{\rho_f}} \PC{\hat{\mathbf{G}}^{'*} \hat{\mathbf{G}}^{'T} + \frac{K \sigma_{w}^2}{E_{tr}}  \mathbf{I}_M}^{-1} \hat{\mathbf{G}}^{'*} \mathbf{N}^{-1}\\
& \qquad \mathbf{N}_{\text{MMSE}} \ \mathbf{s}  + w_k \\
&= \underbrace{\sqrt{\rho_f} \hat{\mathbf{g}}^{'T}_k \frac{f_{\text{MMSE}}}{\sqrt{\rho_f}} \PC{\hat{\mathbf{G}}^{'*} \hat{\mathbf{G}}^{'T} + \frac{K \sigma_{w}^2}{E_{tr}}  \mathbf{I}_M}^{-1} \hat{\mathbf{G}}^{'*} \mathbf{N}^{-1} \mathbf{N}_{\text{MMSE}} \mathbf{s}}_{\text{desired signal + interference}} + \\
&\underbrace{\sqrt{\rho_f} \ \tilde{\mathbf{g}}^{'T}_k \frac{f_{\text{MMSE}}}{\sqrt{\rho_f}} \PC{\hat{\mathbf{G}}^{'*} \hat{\mathbf{G}}^{'T} + \frac{K \sigma_{w}^2}{E_{tr}}  \mathbf{I}_M}^{-1} \hat{\mathbf{G}}^{'*}  \mathbf{N}^{-1} \mathbf{N}_{\text{MMSE}} \ \mathbf{s}}_{\text{CSI error}}\\
& \qquad + \ w_k,
\end{split}
\end{equation}
where $\hat{\mathbf{g}}^{'}_k = [\hat{g}^{'}_{1,k},\dots,\hat{g}^{'}_{M,k}]^T$ is the CSI vector for user $k$ and $\tilde{\mathbf{g}}^{'}_k = [\tilde{g}^{'}_{1,k},\dots,\tilde{g}^{'}_{M,k}]^T$ is the CSI error vector for user $k$\textcolor{r}{, both after APS} .

\textcolor{r}{Assuming uncorrelated Gaussian noise, the achievable rate of user $k$ with the proposed iterative MMSE precoder is given by}
\begin{equation}
R_{k,\text{MMSE}} = \log_2 (1+ \text{SINR}_{k,\text{MMSE}}).
\end{equation}
The sum-rate is then given by
\begin{equation}
R_{\text{MMSE}} = \sum_{k=1}^K \log_2 (1+ \text{SINR}_{k,\text{MMSE}}),
\end{equation}
where the SINR reads as
\begin{equation}
\text{SINR}_{k,\text{MMSE}} = \frac{\mathbb{E} \PR{|A_1|^2}}{\sigma_w^2 + \sum_{i=1,i \neq k}^{K} \mathbb{E} \PR{|A_{2,i}|^2}+\mathbb{E} \PR{|A_3|^2}}.
\end{equation}
\textcolor{r}{Note that the considered SINR is different to the SINR expression used in \cite{Nayebi2017}, which relies on ZF precoding}.
In the expression above, the quantity
\begin{equation}
    A_1 = \sqrt{\rho_f} \hat{\mathbf{g}}^{'T}_k \mathbf{p}_{k} \sqrt{\eta_k} s_k,
\end{equation}
is the desired signal, the parameter
\begin{equation}
    A_{2,i} = \sqrt{\rho_f} \hat{\mathbf{g}}^{'T}_k \mathbf{p}_{i} \sqrt{\eta_i} s_i,
\end{equation}
is the interference caused by user $i$ and
\begin{equation}
    A_3 = \sqrt{\rho_f} \tilde{\mathbf{g}}^{'T}_k \mathbf{P}_{\text{MMSE}} \mathbf{N}_{\text{MMSE}} \mathbf{s}
\end{equation}
refers to CSI error.
The mean-square values of $A_1$, $A_{2,i}$ and $A_3$ are computed as follows:
\begin{equation}
\begin{split}\label{mean_square_1}
\mathbb{E} \PR{|A_1|^2} &=\textcolor{r}{ \mathbb{E} \PR{\PC{\sqrt{\rho_f}\hat{\mathbf{g}}^{'T}_k \mathbf{p}_{k} \sqrt{\eta_k} s_k}^{*} \PC{\sqrt{\rho_f}\hat{\mathbf{g}}^{'T}_k \mathbf{p}_{k} \sqrt{\eta_k} s_k}}}\\
&\textcolor{r}{= \rho_f \eta_k \psi_k }\\
\end{split}
\end{equation}
\begin{equation}
\begin{split}\label{mean_square_2}
\mathbb{E} \PR{|A_{2,i}|^2} &= \textcolor{r}{\mathbb{E} \PR{\PC{\sqrt{\rho_f}\hat{\mathbf{g}}^{'T}_k \mathbf{p}_{i} \sqrt{\eta_i} s_i}^{*} \PC{\sqrt{\rho_f}\hat{\mathbf{g}}^{'T}_k \mathbf{p}_{i} \sqrt{\eta_i} s_i}}}\\
&\textcolor{r}{= \rho_f \eta_i \phi_{k,i}}\\
\end{split}
\end{equation}
\begin{equation}
\begin{split}\label{mean_square_3}
\mathbb{E} \PR{|A_3|^2} &= \textcolor{r}{\mathbb{E} \PR{\left\lvert\sqrt{\rho_f}\tilde{\mathbf{g}}^{'T}_k \mathbf{P}_{\text{MMSE}} \mathbf{N}_{\text{MMSE}} \mathbf{s}\right\rvert^2}}\\
&\textcolor{r}{= \rho_f \sum_{i=1}^K \eta_i \gamma_{k,i}}
\end{split}
\end{equation}
In the expressions above, we have
\begin{equation}
    \psi_k = \mathbf{p}_{k}^H \hat{\mathbf{g}}^{'*}_k \hat{\mathbf{g}}^{'T}_k \mathbf{p}_{k}, \text{ for } k= 1,\dots, K,
\end{equation}
\begin{equation}
   \phi_{k,i} = \mathbf{p}_{i}^H \hat{\mathbf{g}}^{'*}_k \hat{\mathbf{g}}^{'T}_k \mathbf{p}_{i},\text{ for } i \neq k, i = 1,\dots,K,
\end{equation}
\begin{equation}
  \boldsymbol{\gamma}_{k} = \text{diag} \chav{\mathbf{P}_{\text{MMSE}}^H\mathbb{E} \PR{\tilde{\mathbf{g}}^{'*}_k \tilde{\mathbf{g}}^{'T}_k} \mathbf{P}_{\text{MMSE}}},
\end{equation}
where $\mathbf{p}_{k} = \PR{p_{1,k},\dots,p_{M,k}}^T$ is the column $k$ of matrix $\mathbf{P}_{\text{MMSE}}$, $\psi_k$ is the $k$th element of vector $\boldsymbol{\psi}$, $\phi_{k,i}$ is the $i$th element of vector $\boldsymbol{\phi}_k$, $\gamma_{k,i}$ is the $i$th element of vector $\boldsymbol{\gamma}_{k}$, and  $\mathbb{E} \PR{\tilde{\mathbf{g}}^{'*}_k \tilde{\mathbf{g}}^{'T}_k}$ is a diagonal matrix with $\PC{(1-\textcolor{r}{n_{m,k}})\beta^{'}_{mk}}$ on its $m$th diagonal element.

By substituting \eqref{mean_square_1}, \eqref{mean_square_2} and \eqref{mean_square_3} in the $\text{SINR}_{k,\text{MMSE}}$ expression we get
\begin{equation} \label{sinr_mmse_opt}
\text{SINR}_{k,\text{MMSE}} = \frac{\rho_f \eta_k \psi_k}{\sigma_{w}^2 + \rho_f \sum_{i=1,i \neq k}^{K} \eta_i \phi_{k,i} + \rho_f \sum_{i=1}^K \eta_i \gamma_{k,i}}.
\end{equation}
Note that in $\text{SINR}_{k,\text{MMSE}}\PC{\boldsymbol{\eta}}$, the numerator and denominator are linear functions of $\boldsymbol{\eta}$. Consequently, $\text{SINR}_{k,\text{MMSE}}\PC{\boldsymbol{\eta}}$ is a quasilinear function, enabling us to use the bisection method \cite{Boyd2019}. The proof of expression \eqref{sinr_mmse_opt} is detailed in Appendix~\ref{sec_appendix}.

\subsection{Computational Complexity}
\label{comp_complexity}
Here, we evaluate the computational complexity of the proposed and existing techniques \cite{Watkins2014}.

The overall complexity of the proposed iterative LS-APS with MMSE
precoding and OPA is comparable to the existing ZF precoder and is
higher than that of the CB precoder from \cite{Ngo2017,Nayebi2017}.
\textcolor{r}{Depending on the number of iterations of the bisection
method, $T_{\text{OPA}}$, the complexity of the OPA algorithm may
prevail, or the one of the MMSE and ZF precoder, if $M^3 >
T_{\text{OPA}}K^{3.5}$. The same can be said regarding the APA
algorithm, depending on $M^3 > T_{\text{APA}}MK^2$, or the
opposite.} When evaluating the impact of APS on precoding and SINR
computation, there is a reduction in computational complexity from
$\mathcal{O} (M^3)$ to $\mathcal{O}((SN)^3)$ and from
$\mathcal{O}(M^2K^2)$ to $\mathcal{O}((SN)^2K^2)$. For the power
allocation techniques proposed, APS only affects the proposed
\textcolor{r}{APA} and UPA algorithms, with a reduction in
complexity from $\mathcal{O}(T_{\text{APA}}MK^2)$ to
$\mathcal{O}(T_{\text{APA}}SNK^2)$ and from $\mathcal{O} (MK^2)$ to
$\mathcal{O}(SNK^2)$, \textcolor{r}{respectively}, where $M = LN >
SN$. For the precoders with ES-APS, we notice that the complexity is
$\mathcal{O}\PC{L!}$, which is very costly when compared to LS-APS.
Thus, for ES-APS we only consider a very small system with $L=5$
single-antenna APs.


\section {Simulations}
In this section, we assess the proposed and existing techniques using simulations. The proposed APS is first applied, followed by MMSE precoder, which is initially calculated with $\mathbf{N} = \mathbf{I}_K$, and then power allocation is performed. With the corresponding power coefficients, the precoder is recomputed and the final power allocation is performed, completing two iterations in total. In all experiments, we performed 120 channel realizations and assumed $\sigma_s^2 = 1$. \textcolor{r}{Moreover, we consider that $n_{m,k} = n$, which means that all channel coefficients are estimated with the same accuracy.}

We consider $L$ APs with $N$ antenna elements, a total of $M = LN$ antennas, and $K$ single-antenna users uniformly distributed at random within an area of $1 \text{ km}^2$. The LS coefficients from \eqref{def_g_mk} are modeled by
\begin{equation}
    \beta_{m,k} = \text{PL}_{m,k} \cdot 10^{\frac{\sigma_{sh} z_{m,k}}{10}},
\end{equation}
where $\text{PL}_{m,k}$ is the path loss and $10^{\frac{\sigma_{sh} z_{m,k}}{10}}$ refers to the shadow fading with standard deviation $\sigma_{sh} = 8$ dB and $z_{m,k} \sim \mathcal{N} \PC{0,1}$.
The path loss is based on a three-slope model \cite{Tang2001}, in dB, defined as
\begin{footnotesize}
\begin{equation}
\text{PL}_{m,k} =
\begin{cases}
    -L-35 \log_{10} \PC{d_{m,k}}, \text{ if } d_{m,k} > d_1\\
    -L-15\log_{10} \PC{d_1} - 20\log_{10}\PC{d_{m,k}},\\
    \qquad \qquad \qquad \text{ if } d_{0} < d_{m,k} \leq d_1\\
    -L-15\log_{10} \PC{d_1} - 20\log_{10}\PC{d_0} , \text{ if } d_{m,k} \leq d_0\\
    \end{cases}
\end{equation}
\end{footnotesize}
where
\begin{equation}\begin{split}
L &\triangleq 46.3 + 33.9 \log_{10} \PC{\textcolor{r}{f_{\text{freq}}}} - 13.82 \log_{10} \PC{h_{\text{AP}}}\\
& \qquad - \PC{1.1 \log_{10} \PC{\textcolor{r}{f_{\text{freq}}}} - 0.7}h_{\text{u}} + \PC{1.56 \log_{10} \PC{\textcolor{r}{f_{\text{freq}}}} - 0.8},
\end{split}
\end{equation}
$d_{m,k}$ is the distance between the $m$th antenna element and the $k$th user, $d_1 = 50$ m, $d_0 = 10$ m, $\textcolor{r}{f_{\text{freq}}} = 1900$ MHz is the carrier frequency in MHz, $h_{\text{AP}} = 15$ m is the AP antenna height in meters and $h_{\text{u}} = 1.65$ m is the user antenna height in meters, as in \cite{Ngo2017}. When $d_{m,k} \leq d_1$ there is no shadowing.

We consider strong path loss, which is typical of Cell-Free Massive MIMO systems, and define $\rho_f$ based on the signal-to-noise ratio (SNR) given by \cite{Brown2012}
\begin{equation} \label{rho_f_exp}
    \rho_f = \frac{\text{SNR} \cdot \text{tr}\PC{\mathbf{C}_w}}{\mathbb{E} [ || \hat{\mathbf{G}}||_F^2 ]} = \frac{\text{SNR} \cdot K \sigma_{w}^2}{\text{tr} (\hat{\mathbf{G}} \hat{\mathbf{G}}^H)},
\end{equation}
where
\begin{equation}
    \sigma_{w}^2 = T_0 \times k_B \times B \times NF \text{(W)},
\end{equation}
$T_0 = 290$ (Kelvin) is the noise temperature, $k_B = 1.381 \times 10^{-23}$ (Joule per Kelvin) is the Boltzmann constant, $B = 20$ MHz is the bandwidth and $NF = 9$ dB is the noise figure. Therefore, the SNR expression is
\begin{equation} \label{snr_exp}
    \text{SNR} = \frac{\rho_f \mathbb{E} [ || \hat{\mathbf{G}}||_F^2 ]}{\text{tr}\PC{\mathbf{C}_w}} = \frac{\rho_f \text{tr} (\hat{\mathbf{G}} \hat{\mathbf{G}}^H)}{K \sigma_{w}^2}.
\end{equation}
Since we are comparing different precoding designs in this work, an abstract SNR expression is being considered. Note that the expression does not take in account the beamforming gain. Thus, the SINR can be higher than the SNR in the numerical results \cite{Tse2005}.

In the experiments we will present next, we combined different precoding, power allocation and APS techniques. Therefore, when describing a technique we will use the following notation:
\begin{itemize}
    \item Precoding + Power Allocation + APS
\end{itemize}
For each category, we have the following methods:
\begin{itemize}
    \item Precoding: CB, ZF and the Proposed MMSE
    \item Power Allocation: OPA, APA and UPA
    \item APS: ES-APS and LS-APS
\end{itemize}

\begin{figure}[!ht]
\centering
\includegraphics[width = 1\columnwidth]{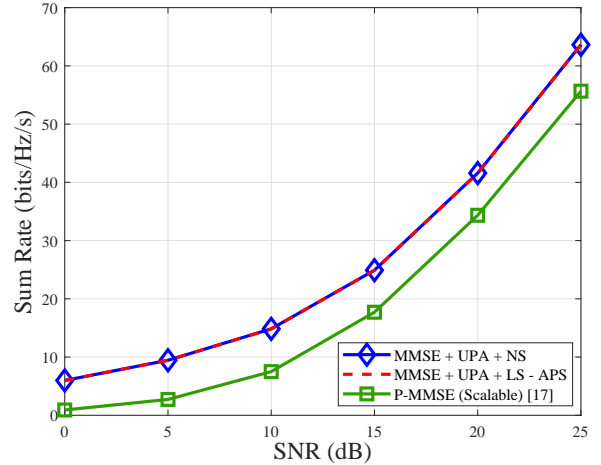}
\caption{\textcolor{r}{Sum-Rate vs. SNR with $M$ = 128, $K = 16$, $n = 1$ and $150$ channel realizations. }}\label{iterative_vs_scalable}
\end{figure}

\textcolor{r}{In the first experiment, we present a comparison of the proposed techniques and the centralized scheme from \cite{Bjornson2020a}. In this comparison, we consider MMSE+UPA with NS and LS-APS. From \cite{Bjornson2020a}, we considered the P-MMSE technique and the developed clustering method, \textcolor{red}{which} implies AP\textcolor{red}{S}. The precoding scheme is actually a centralized MMSE precoder combined with \textcolor{red}{a} heuristic power allocation strategy. For this simulation, we employ the same distance matrix, with coefficients $d_{m,k}$, so that each method is able to calculate its own AP\textcolor{red}{S} and precoding vectors. We perform 150 channel realizations under perfect CSI conditions. Additionally, we consider in both scenarios an SNR variation between 0 and 25 dB, which is described by \eqref{snr_exp}, given $\sigma_w^2 = 1$. As shown in Fig. \ref{iterative_vs_scalable}, gains up to 6 dB can be obtained by the MMSE + UPA schemes over the P-MMSE technique. For a sum-rate equal to 20, differences around 3 dB can be observed when evaluating the compared techniques.}

\begin{figure}[!ht]
\centering
\includegraphics[width = 1\columnwidth]{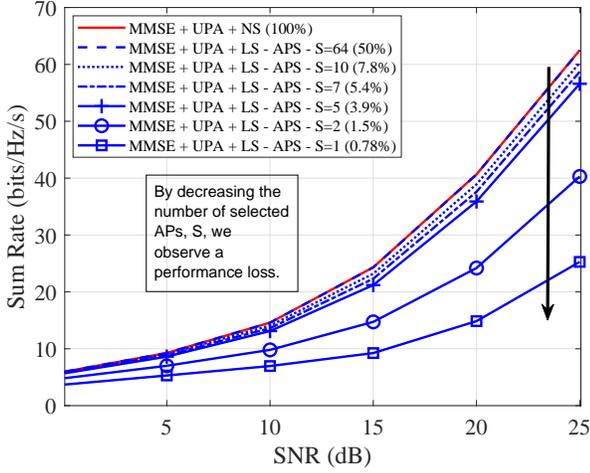}
\caption{\textcolor{red}{Sum-Rate vs. SNR with $L$ = 128, $N$ = 1, $K = 16$, $n = 1$, $120$ channel realizations and $E_{tr} = M \rho_f$. }}\label{sum_rate_selected_aps}
\end{figure}

{Second, we demonstrate the effects of APS in sum-rate performance.
With the decrease in the number of selected APs, a performance
degradation is observed. In Fig.~\ref{sum_rate_selected_aps}, we
notice that with 50\% of selected APs, the performance is comparable
to the case with no selection (NS). However, when less than 10\% of
the available APs are selected for each user, a significant
performance degradation can be perceived. This phenomenon occurs
because the channel matrices are sparse, $\mathbf{G},
\hat{\mathbf{G}}, \tilde{\mathbf{G}}$, which means that only the APs
close to he user are efficiently transmitting signals to it. In
combination with the per antenna power constraint one can conclude
that some channels do not contribute to the received signal and
therefore can be discarded through an APS scheme. Although a weak
channel does not necessarily mean a small contribution to the
received signal, its association with the per antenna power
constraint leads to this effect. When selecting half of the
available APs, most of the available benefits are obtained since the
remaining channels contribute substantially less to performance.
Nevertheless, when less APs are selected, a larger percentage of
links that significantly contribute to performance improvement can
be discarded, leading to performance degradation.}

\begin{figure}[!ht]
\centering
\includegraphics[width=1\columnwidth]{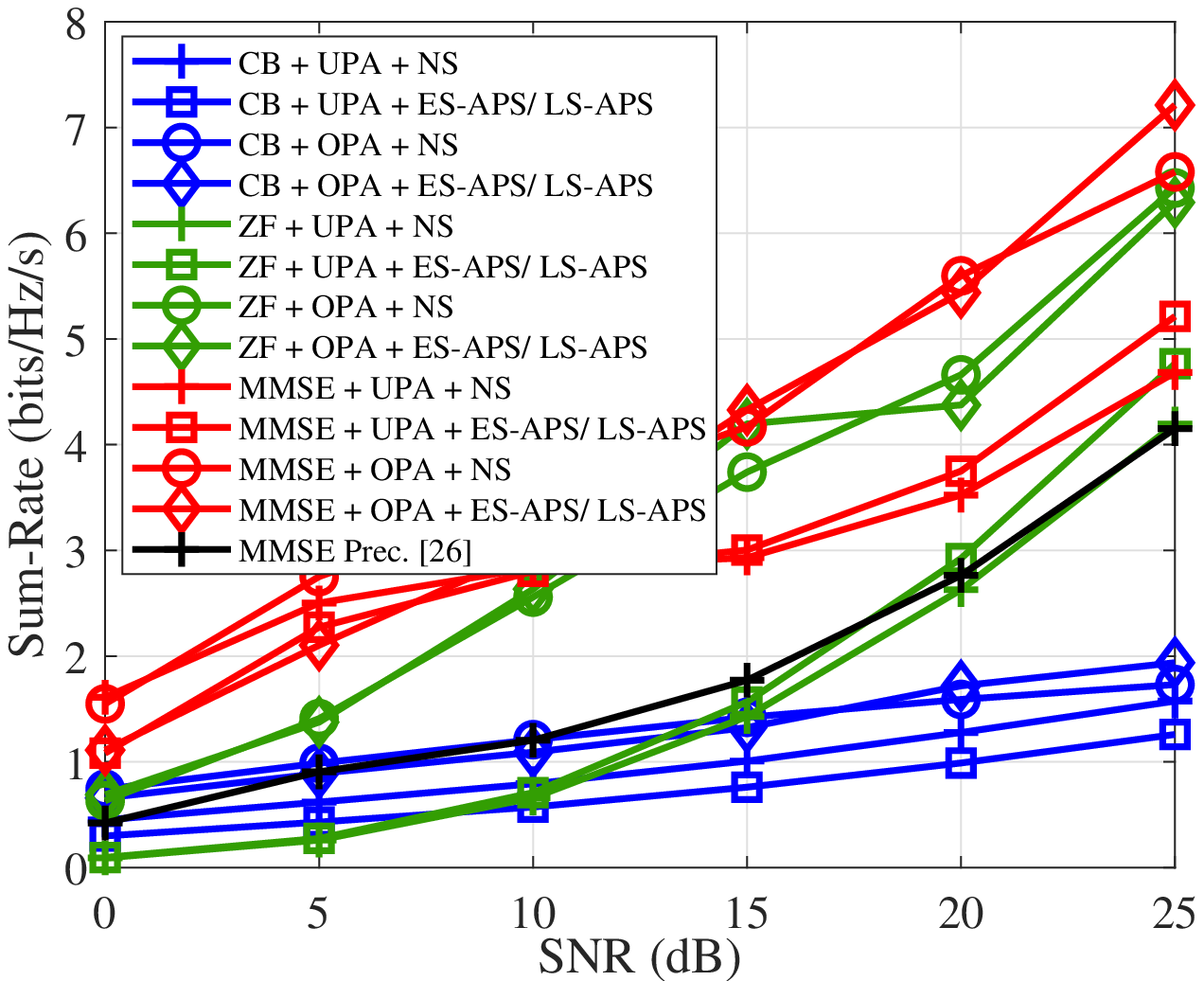}
\caption{Sum-Rate vs. SNR with $L$ = 5, N$ = 1$, $S = 3$, $K = 2$, $n = 0.99$, $120$ channel realizations and $E_{tr} = M\rho_f$.} \label{Sum_Rate_SB_OPT_OPA}
\end{figure}

\begin{figure}[!ht]
\centering
\includegraphics[width=1\columnwidth]{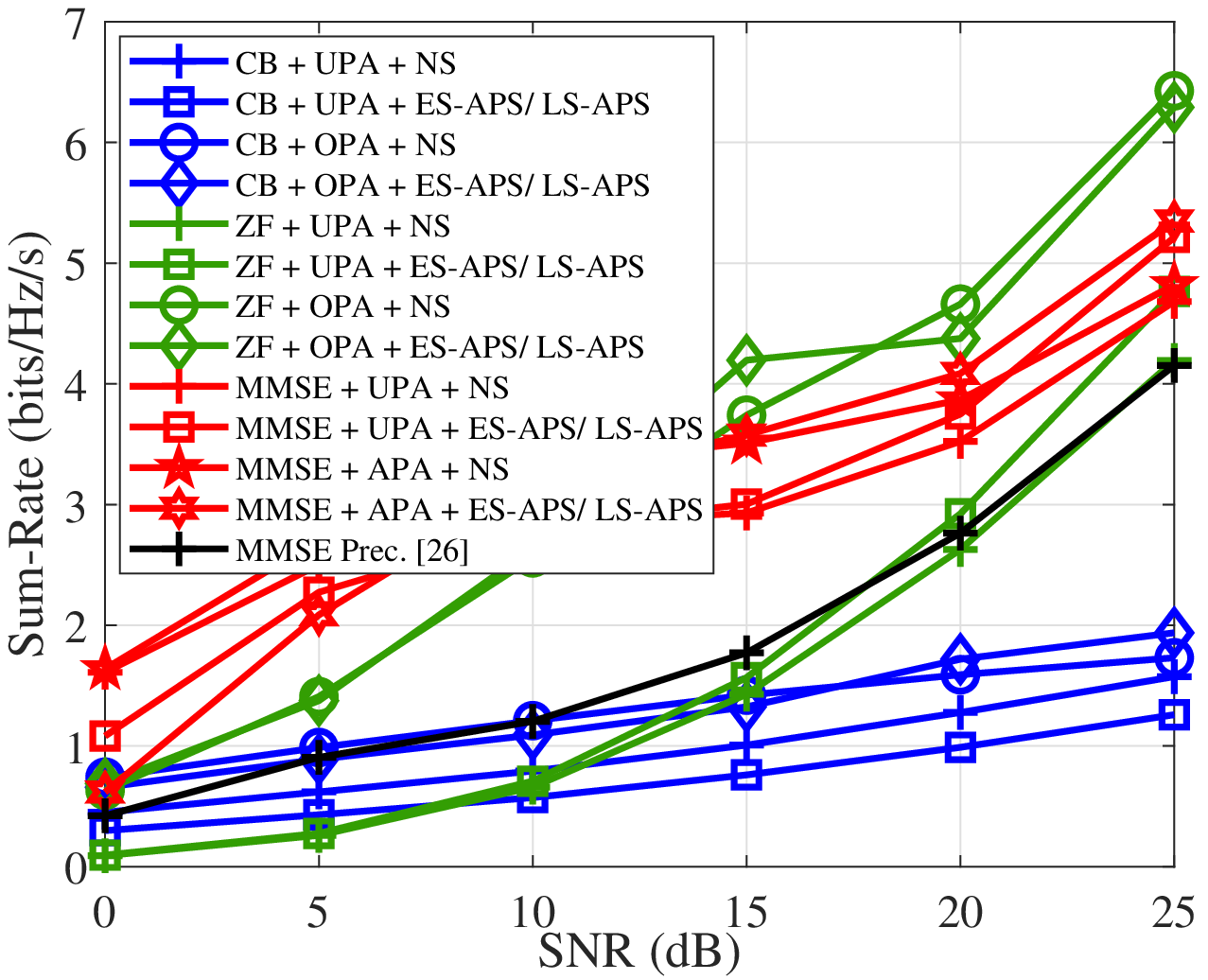}
\caption{Sum-Rate vs. SNR with $L$ = 5, N$ = 1$, $S = 3$, $K = 2$, $n = 0.99$, $120$ channel realizations and $E_{tr} = M\rho_f$.} \label{Sum_Rate_SB_OPT_APA}
\end{figure}

In Fig.~\ref{Sum_Rate_SB_OPT_OPA}, we compare the proposed iterative
MMSE precoder with OPA and UPA with the CB and ZF precoders from
\cite{Ngo2017,Nayebi2017} in terms of sum-rate vs. SNR. In the first
case, both ES-APS and LS-APS are compared to \textcolor{red}{NS}.
Since ES-APS has high computational complexity, a very small system
is considered in Fig.~\ref{Sum_Rate_SB_OPT_OPA} and
Fig.~\ref{Sum_Rate_SB_OPT_APA} with $L=5$ single-antenna APs, $S =
3$ selected APs and $K=2$ users only. In
Fig.~\ref{Sum_Rate_SB_OPT_APA}, we explore the same scenario, but
instead of looking at OPA, we consider APA.

As shown in Fig.~\ref{Sum_Rate_SB_OPT_OPA}, MMSE + OPA is the scheme
with the best performance. It is also visible that MMSE + UPA
achieves higher rates than ZF + OPA for lower SNR values and it is
better than ZF + UPA in the whole experiment. Additionally, we can
see that the application of ES-APS/LS-APS generates comparable or
even improved results for MMSE, ZF and CB + OPA. In the case of CB +
UPA, there is a small degradation in performance when applying {APS}
techniques. We remark that both selection schemes are comparable and
are shown together for aesthetic purposes. {The reason for this is
that the cell-free channel matrix is sparse, which means that most
of the APs far from the user are not efficiently transmitting
signals to it. Therefore, by APS methods, also called clustering
methods, we discard unnecessary transmissions. By using the LS-APS
technique, which chooses the strongest channels to transmit, we can
approach the optimal technique, which tests all arrangements.} As a
result, we conclude that the suboptimal scheme presented is an
effective replacement for the optimal technique but with lower
computational complexity.

Fig.~\ref{Sum_Rate_SB_OPT_APA} provide us some insight on the performance of the MMSE + APA scheme when compared to MMSE + UPA, and ZF. The uniform and adaptive technique\textcolor{r}{s} have a better performance than ZF + OPA for lower values of SNR. Moreover, we can see that the MMSE+APA scheme performs better than MMSE+UPA. In all combined techniques, OPA performs better than UPA and APA. In both figures, the MMSE precoder from \cite{Joham2005} shows a degraded performance due to its lack of appropriate power allocation.

\begin{figure}[!ht]
\centering
\includegraphics[width=1\columnwidth]{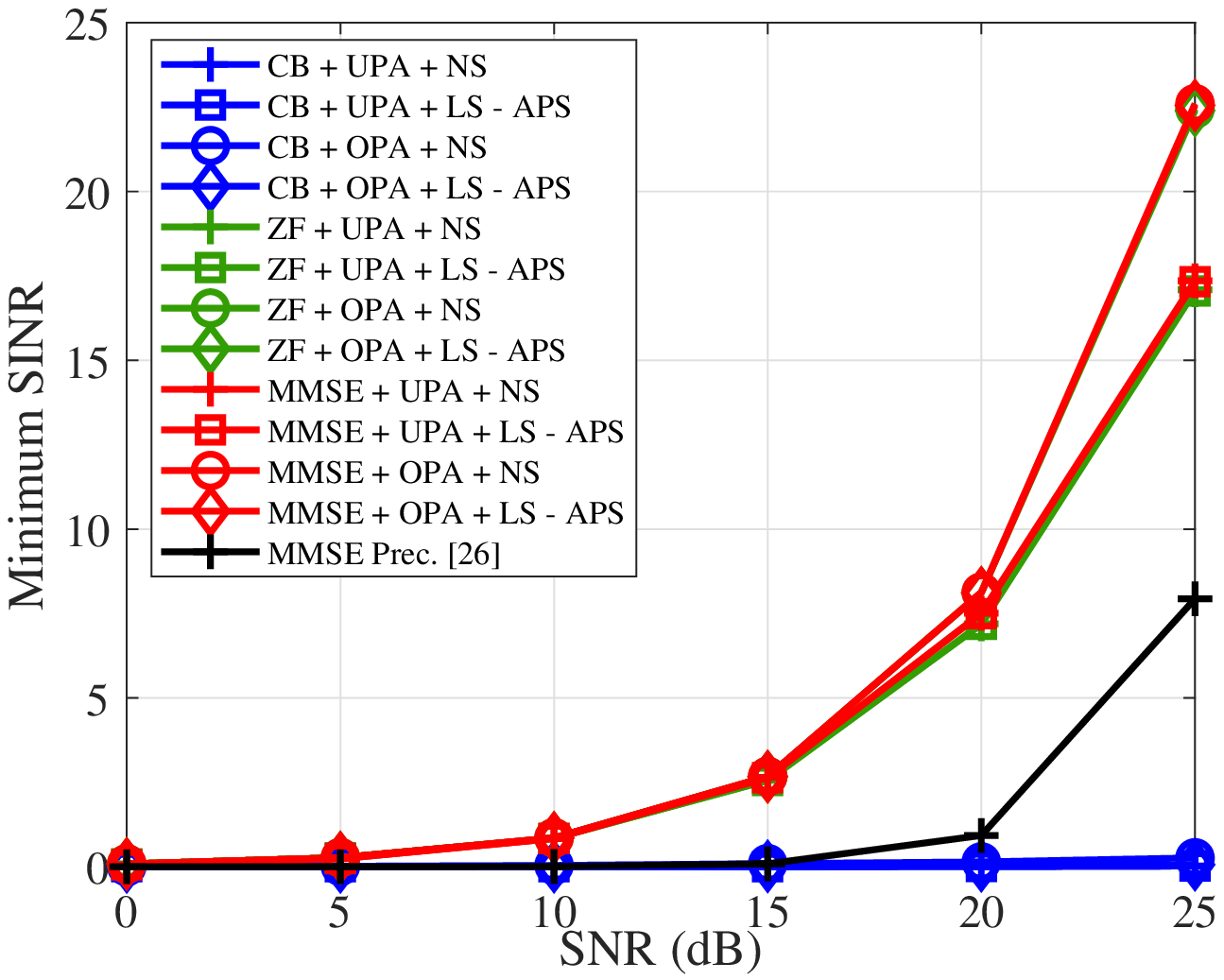}
\caption{Minimum SINR vs. SNR with $L$ = 128, N$ = 1$, $S = 64$, $K = 16$, $n = 0.99$, $120$ channel realizations and $E_{tr} = M\rho_f$.} \label{Min_Rate_SB_SUB_OPA}
\end{figure}

\begin{figure}[!ht]
\centering
\includegraphics[width=1\columnwidth]{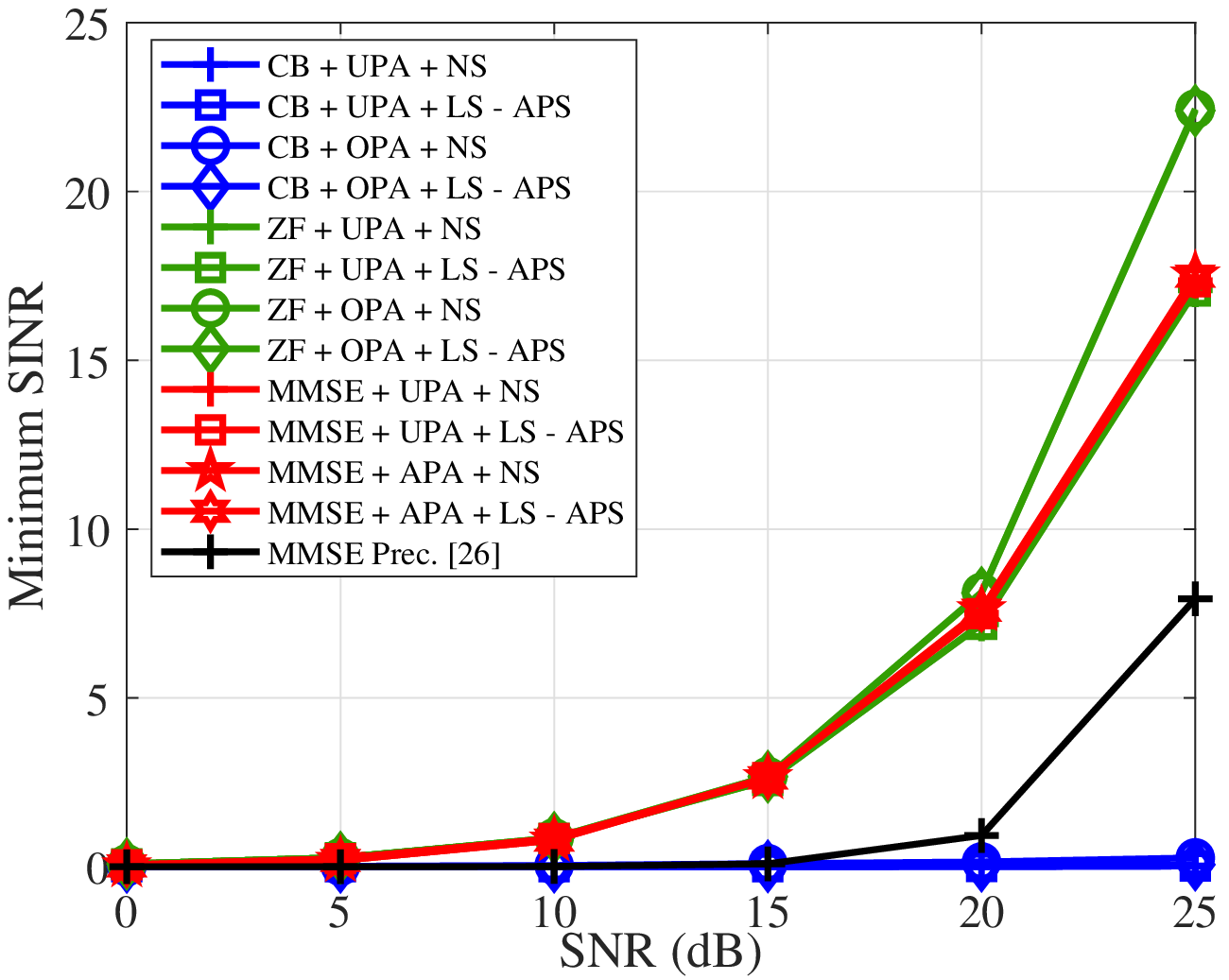}
\caption{Minimum SINR vs. SNR with $L$ = 128, N$ = 1$, $S = 64$, $K = 16$, $n = 0.99$, $120$ channel realizations and $E_{tr} = M\rho_f$.} \label{Min_Rate_SB_SUB_APA}
\end{figure}

\begin{figure}[!ht]
\centering
\includegraphics[width=1\columnwidth]{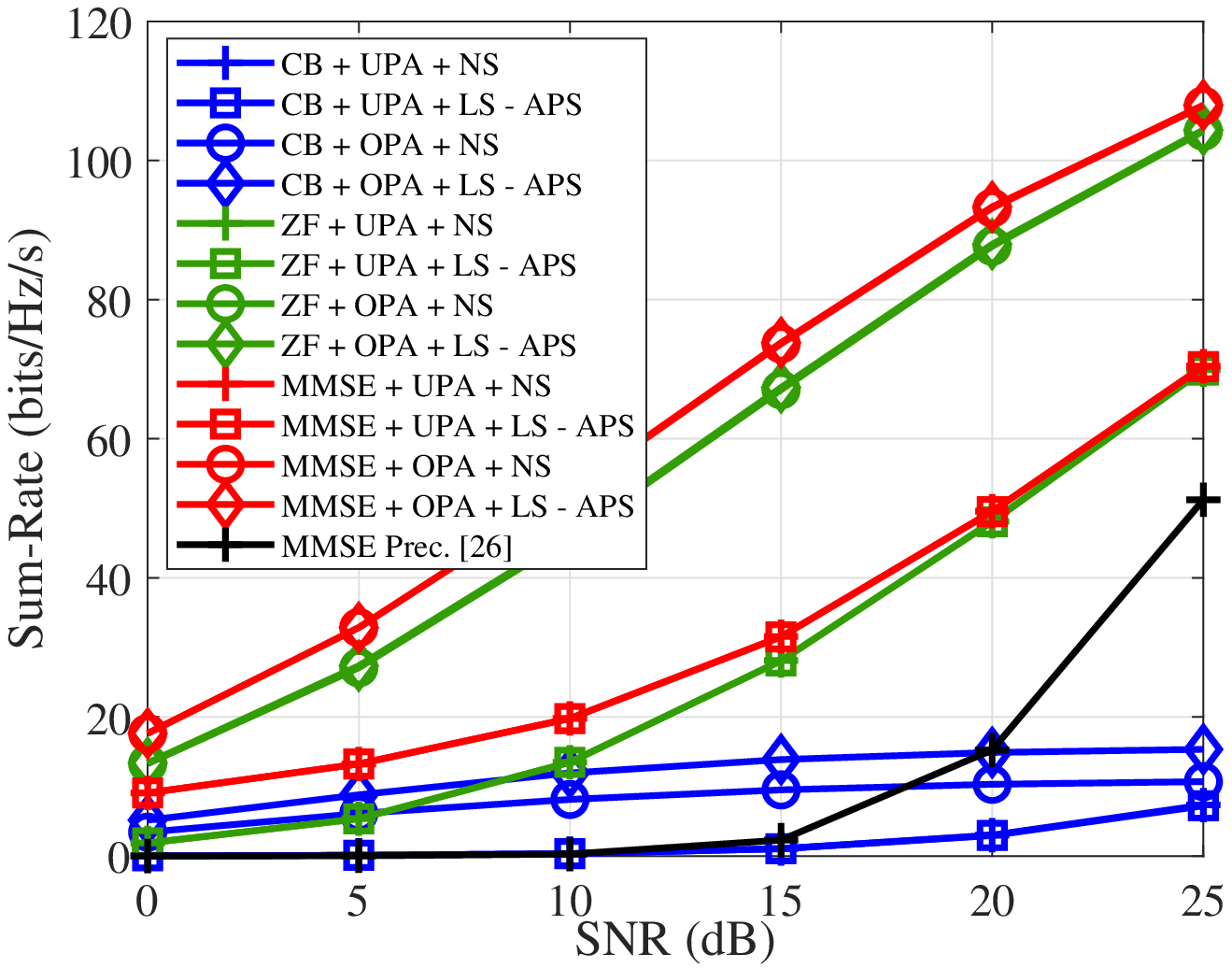}
\caption{Sum-Rate vs. SNR with $L$ = 128, N$ = 1$, $S = 64$, $K = 16$, $n = 0.99$, $120$ channel realizations and $E_{tr} = M\rho_f$.} \label{Sum_Rate_SB_SUB_OPA}
\end{figure}

\begin{figure}[!ht]
\centering
\includegraphics[width=1\columnwidth]{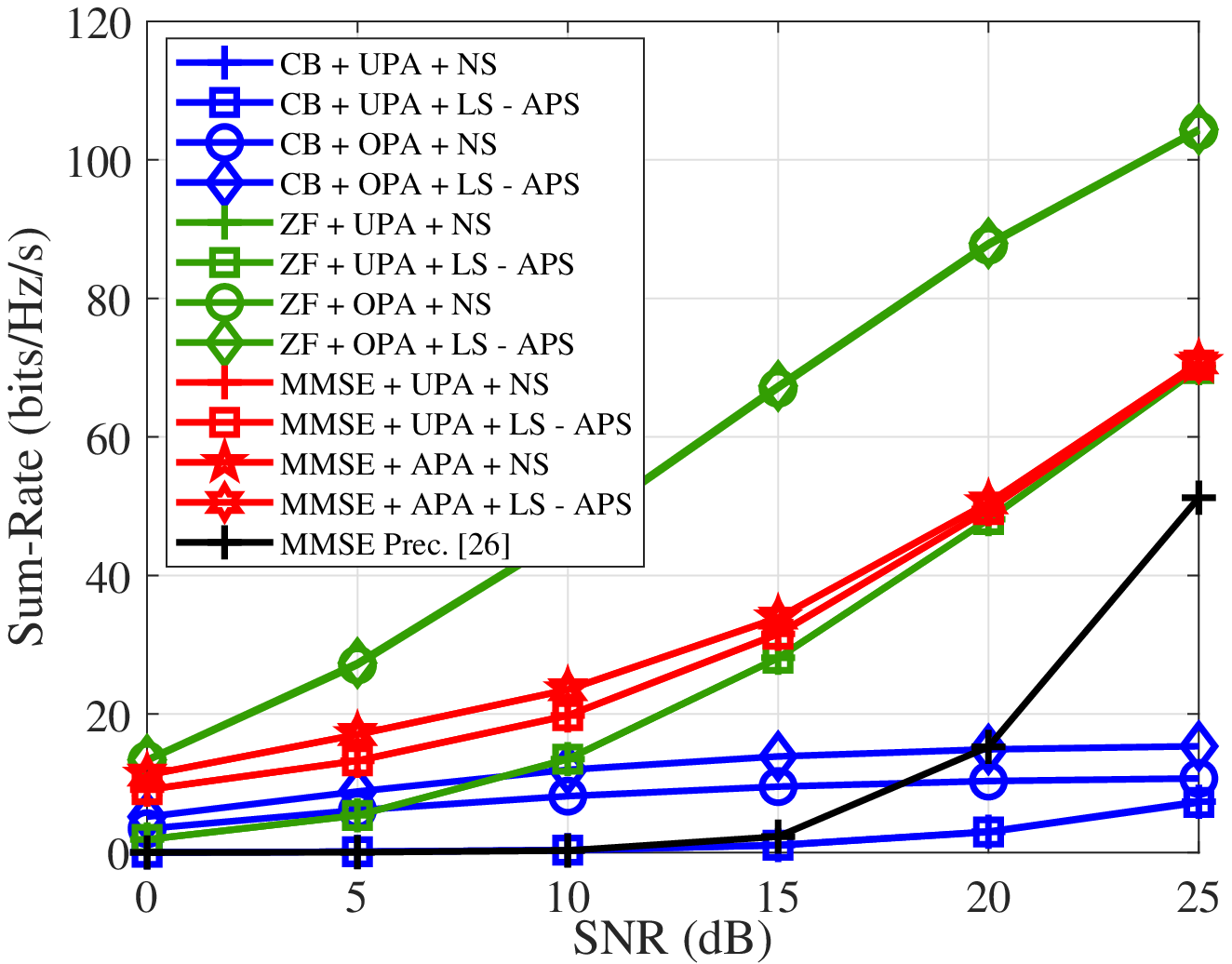}
\caption{Sum-Rate vs. SNR with $L$ = 128, N$ = 1$, $S = 64$, $K = 16$, $n = 0.99$, $120$ channel realizations and $E_{tr} = M\rho_f$.} \label{Sum_Rate_SB_SUB_APA}
\end{figure}

In the second experiment, we explore LS-APS in a large system with $L = 128$ single-antenna APs, $S = 64$ selected APs and $K = 16$ users, in terms of minimum SINR and sum-rate. As we did previously, in Fig.~\ref{Min_Rate_SB_SUB_OPA} and Fig.~\ref{Sum_Rate_SB_SUB_OPA} we compare the CB and ZF precoders from \cite{Ngo2017,Nayebi2017} with OPA and UPA and in Fig.~\ref{Min_Rate_SB_SUB_APA} and Fig.~\ref{Sum_Rate_SB_SUB_APA} we substitute OPA for APA.

The results in Fig.~\ref{Min_Rate_SB_SUB_OPA} and Fig.~\ref{Min_Rate_SB_SUB_APA} validate the functionality of the max-min fairness power allocation algorithm, where our main objective was that the OPA algorithms have at least the same minimum SINR \textcolor{red}{as} UPA ones, if not higher. \textcolor{r}{As shown in Fig.~\ref{Min_Rate_SB_SUB_OPA}, in higher SINR values, we can see the considerable improvement provided by the OPA algorithm, when compared to UPA. On the other hand, in Fig.~\ref{Min_Rate_SB_SUB_APA}~, we study APA instead of OPA and we are able to observe that its performance is comparable to the one of the UPA scheme. We can conclude that the OPA technique improves the minimum SINR, as opposed to APA and UPA, where the goal is to improve the overall performance of the system.}

In Fig.~\ref{Sum_Rate_SB_SUB_OPA}, the MMSE + OPA scheme has the best performance compared to the other schemes. Moreover, the MMSE + UPA technique achieves higher rates than ZF + UPA. We also note that performance is maintained when applying LS-APS for large systems, except for CB + OPA, when performance is improved. For all precoders, OPA provides significantly better rates than UPA. In a larger system, the performance of the precoder from \cite{Joham2005} is also not as good as MMSE +UPA due to its inappropriate design for cell-free systems.

We notice in Fig.~\ref{Sum_Rate_SB_SUB_APA} that MMSE + APA can provide an improvement in performance when compared to MMSE + UPA, in terms of sum-rate. Therefore, it is an attractive solution in comparison with the remaining precoders when combined with UPA.

\begin{figure}[!ht]
\centering
\includegraphics[width = 1\columnwidth]{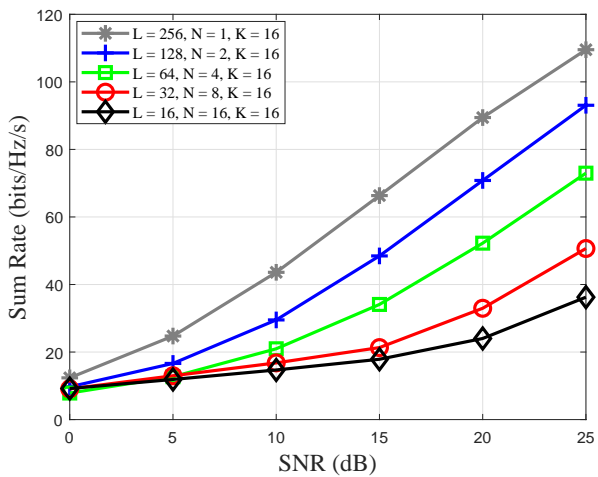}
\caption{\textcolor{r}{Sum-Rate vs. SNR with $M$ = 256, $S = 128$, $K = 16$, $n = 0.99$, $120$ channel realizations and $E_{tr} = M\rho_f$. MMSE+UPA+LS-APS scheme used. }}\label{fixed_numbers}
\end{figure}

\textcolor{r}{In Fig.~\ref{fixed_numbers}, we assess the sum-rate vs. SNR for APs with different numbers of antenna elements. We compare MMSE+UPA+LS-APS under scenarios with a fixed total number of antennas, $M=256$, but with different numbers of antennas per AP. Fig.~\ref{fixed_numbers} verifies that the best sum-rate performance is achieved with a single-antenna AP scenario. As pointed out in \cite{Chen2018}, although adding more antennas to an AP leads to more channel hardening, it also brings losses in macro-diversity, resulting in lower average rates. Therefore, the best settings for the proposed scenarios are presented in Figs.~\ref{Sum_Rate_SB_OPT_OPA} to \ref{Sum_Rate_SB_SUB_APA}.}

The last experiment also considers a large system, but now in terms of BER vs. SNR using multiple-antenna APs. We assume perfect CSI ($n = 1$) and QPSK modulation. We consider LS-APS with an antenna array of $N = 4$ elements each, $L = 24$ APs (total of $M=96$ antennas), $S = 12$ selected APs (total of 48 selected antennas) and $K = 8$ users.

\begin{figure}[!ht]
\centering
\includegraphics[width=1\columnwidth]{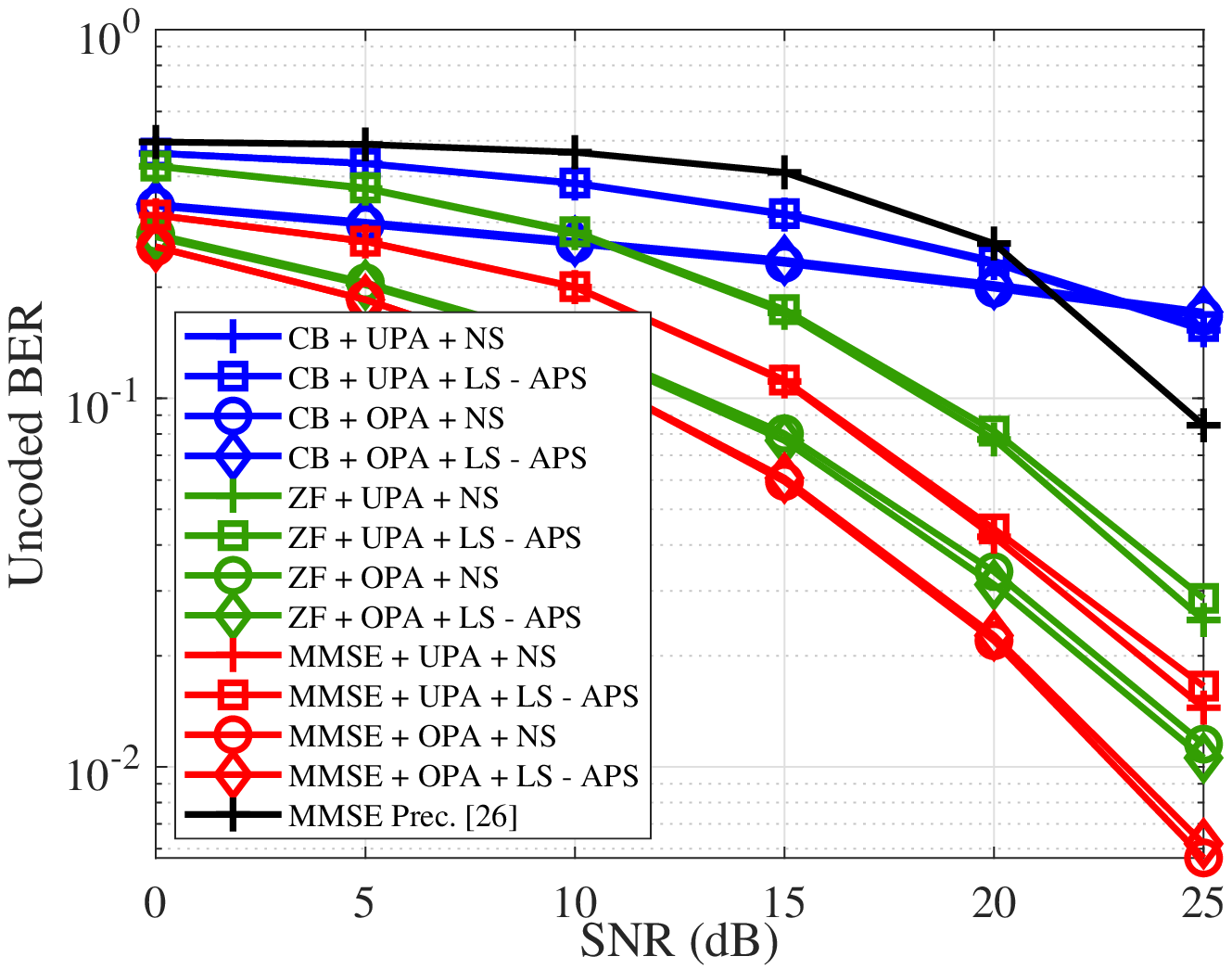}
\caption{BER vs. SNR with $L = 24$, $N=4$, $S = 12$, $K = 8$, \textcolor{r}{$n = 1$}, $120$ channel realizations, $100$ symbols per packet and $E_{tr} = M\rho_f$.}\label{BER_OPA}
\end{figure}

\begin{figure}[!ht]
\centering
\includegraphics[width=1\columnwidth]{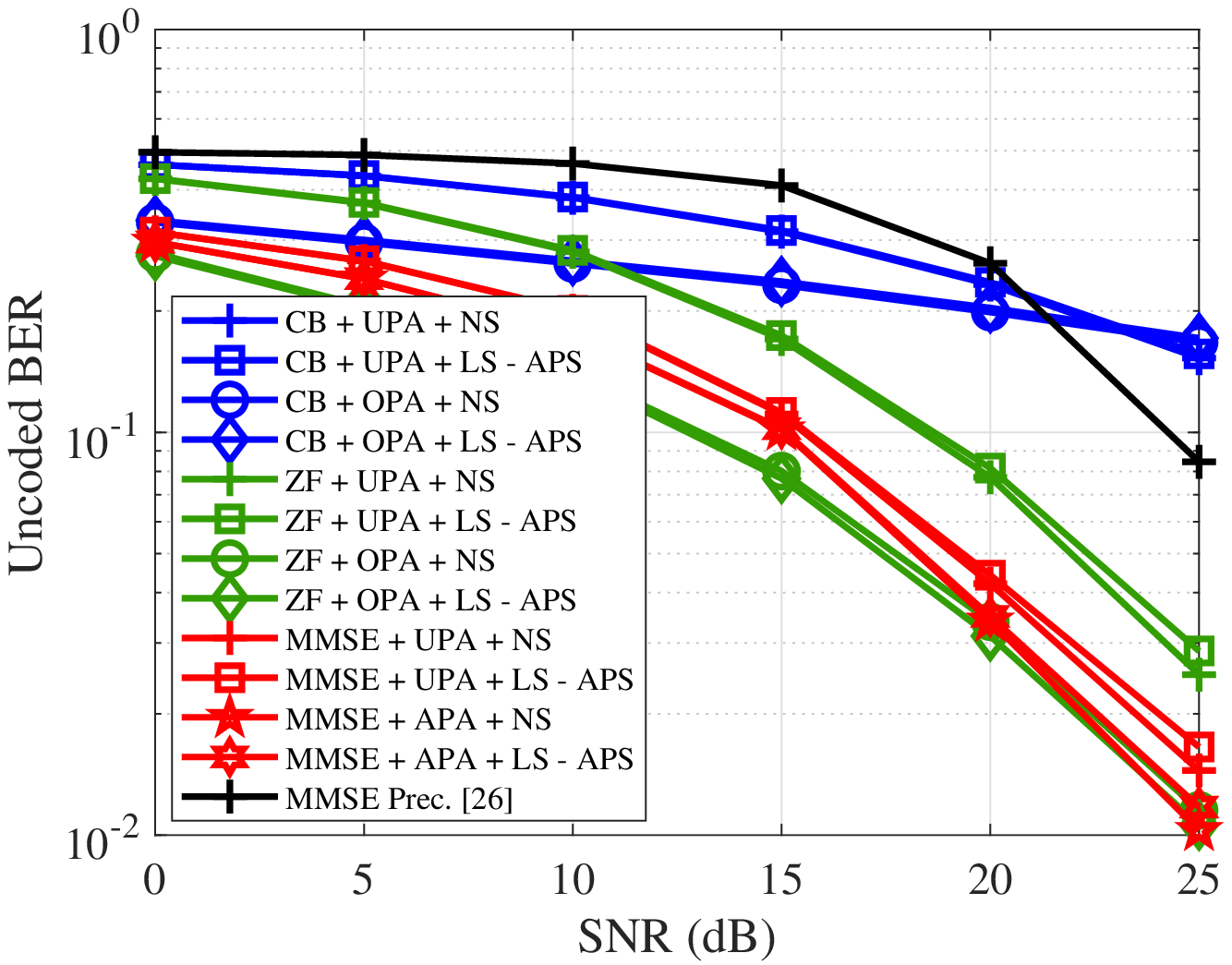}
\caption{BER vs. SNR with $L = 24$, $N=4$, $S = 12$, $K = 8$, \textcolor{r}{$n = 1$}, $120$ channel realizations, $100$ symbols per packet and $E_{tr} = M\rho_f$.}\label{BER_APA}
\end{figure}

As in experiment 2, LS-APS causes no degradation in performance, with the benefit of reducing the computational complexity by half. The insight provided by Fig.~\ref{BER_OPA} and Fig.~\ref{BER_APA} is the same as before. MMSE + OPA has the best performance when compared to other precoders and OPA performs better when applied to all precoders. We also emphasize here that MMSE + APA is a promising solution against UPA and improves performance. Additionally, we remark that in terms of BER, for higher \textcolor{r}{SNR values}, MMSE + APA is comparable to ZF + OPA. Future work might focus on detection and decoding techniques for cell-free networks \cite{deLamare2003,itic,deLamare2008,cai2009,jiomimo,dfcc,deLamare2013,did,rrmser,jidf,saalt,bfidd,1bitidd,aaidd,listmtc,dynmtc,mwc,dynovs,dopeg,memd,vfap}.

\section{Conclusions}

We have presented iterative APS, MMSE precoding and power allocation
techniques for the downlink of a Cell-Free Massive MIMO system with
single and multiple-antenna APs, in the presence of perfect and
imperfect CSI. Two APS schemes were presented, one based on ES,
which takes the experiment to its optimal scenario and the other,
less complex but with comparable performance, based on the largest
LS coefficients. A linear MMSE precoder has been developed by taking
into account a power allocation matrix in its derivation. Then,
three power control algorithms are proposed, OPA, APA and UPA, with
different performances, criteria and computational complexities. We
have then derived sum-rate expressions for the proposed scheme along
with a study of the computational cost of all proposed and existing
techniques. Simulations show that the proposed techniques outperform
existing approaches and can reduce computational cost. The
scalability of the proposed system as well as a robust version of
the proposed schemes will be analysed in future works.


\section*{Appendices}\label{sec_appendix}
\textcolor{r}{In the following the derivation of $\text{SINR}_{k,\text{MMSE}}$ is explained.}
\textcolor{r}{
The received signal by user $k$ is written as
\begin{equation}
\begin{split} \label{eq_y_k_2}
y_k &= \sqrt{\rho_f} \ \mathbf{g}^{T}_k \mathbf{P}_{\text{MMSE}} \ \mathbf{N}_{\text{MMSE}} \ \mathbf{s} + w_k \\
&= \sqrt{\rho_f} \ \PC{\hat{\mathbf{g}}^{'}_k + \tilde{\mathbf{g}}^{'}_k}^T \frac{f_{\text{MMSE}}}{\sqrt{\rho_f}} \PC{\hat{\mathbf{G}}^{'*} \hat{\mathbf{G}}^{'T} + \frac{K \sigma_{w}^2}{E_{tr}}  \mathbf{I}_M}^{-1} \hat{\mathbf{G}}^{'*} \mathbf{N}^{-1}\\
& \qquad \mathbf{N}_{\text{MMSE}} \ \mathbf{s}  + w_k \\
&= \underbrace{\sqrt{\rho_f} \ \hat{\mathbf{g}}^{'T}_k \frac{f_{\text{MMSE}}}{\sqrt{\rho_f}} \PC{\hat{\mathbf{G}}^{'*} \hat{\mathbf{G}}^{'T} + \frac{K \sigma_{w}^2}{E_{tr}}  \mathbf{I}_M}^{-1} \hat{\mathbf{G}}^{'*} \mathbf{N}^{-1} \mathbf{N}_{\text{MMSE}}  \ \mathbf{s}}_{\text{desired signal + interference}} + \\
&\underbrace{\sqrt{\rho_f} \ \tilde{\mathbf{g}}^{'T}_k \frac{f_{\text{MMSE}}}{\sqrt{\rho_f}} \PC{\hat{\mathbf{G}}^{'*} \hat{\mathbf{G}}^{'T} + \frac{K \sigma_{w}^2}{E_{tr}}  \mathbf{I}_M}^{-1} \hat{\mathbf{G}}^{'*}  \mathbf{N}^{-1} \mathbf{N}_{\text{MMSE}} \ \mathbf{s}}_{\text{CSI error}}\\
&+ w_k.
\end{split}
\end{equation}
In \eqref{eq_y_k_2}, we know that channel coefficients, data symbols and noise are mutually independent, which allows us to show that the desired signal, the interference and the channel estimation error are mutually uncorrelated. As stated by \cite{Hassibi2003} and based on the worst case uncorrelated additive noise, the achievable rate is given by $\log_2 (1+ \text{SINR}_{k,\text{MMSE}})$, where
\begin{equation}
\text{SINR}_{k,\text{MMSE}} = \frac{\mathbb{E} \PR{|A_1|^2}}{\sigma_w^2 + \sum_{i=1,i \neq k}^{K} \mathbb{E} \PR{|A_{2,i}|^2}+\mathbb{E} \PR{|A_3|^2}}. \label{sinr_k_mmse_2}
\end{equation}
The quantities in \eqref{sinr_k_mmse_2} are given by
\begin{equation}
    A_1 = \sqrt{\rho_f} \hat{\mathbf{g}}^{'T}_k \mathbf{p}_{k} \sqrt{\eta_k} s_k,
\end{equation}
which is the desired signal,
\begin{equation}
    A_{2,i} = \sqrt{\rho_f} \hat{\mathbf{g}}^{'T}_k \mathbf{p}_{i} \sqrt{\eta_i} s_i,
\end{equation}
which is the interference caused by user $i$ and
\begin{equation}
    A_3 = \sqrt{\rho_f} \tilde{\mathbf{g}}^{'T}_k \mathbf{P}_{\text{MMSE}} \mathbf{N}_{\text{MMSE}} \mathbf{s}
\end{equation}
which refers to CSI error.
The mean-square values of $A_1$, $A_{2,i}$ and $A_3$ can be cast as
\begin{equation}
\begin{split}
\mathbb{E} \PR{|A_1|^2} &= \mathbb{E} \PR{\PC{\sqrt{\rho_f}\hat{\mathbf{g}}^{'T}_k \mathbf{p}_{k} \sqrt{\eta_k} s_k}^{*} \PC{\sqrt{\rho_f}\hat{\mathbf{g}}^{'T}_k \mathbf{p}_{k} \sqrt{\eta_k} s_k}}\\
&= \mathbb{E} \PR{\rho_f \eta_k s_k^{*} \mathbf{p}_{k}^H  \hat{\mathbf{g}}^{'*}_k\hat{\mathbf{g}}^{'T}_k \mathbf{p}_{k} s_k}\\
&= \rho_f \text{tr} \PC{\mathbb{E} \PR{\eta_k \mathbf{p}_{k}^H  \hat{\mathbf{g}}^{'*}_k \hat{\mathbf{g}}^{'T}_k \mathbf{p}_{k} s_k s_k^{*}}}\\
&= \rho_f \text{tr} \PC{\eta_{k}  \mathbf{p}_{k}^H \hat{\mathbf{g}}^{'*}_k \hat{\mathbf{g}}^{'T}_k \mathbf{p}_{k}}\\
&= \rho_f \eta_k \psi_k, \\
\end{split} \label{mean_square_1_2}
\end{equation}
\begin{equation}
\begin{split}
\mathbb{E} \PR{|A_{2,i}|^2} &= \mathbb{E} \PR{\PC{\sqrt{\rho_f}\hat{\mathbf{g}}^{'T}_k \mathbf{p}_{i} \sqrt{\eta_i} s_i}^{*} \PC{\sqrt{\rho_f}\hat{\mathbf{g}}^{'T}_k \mathbf{p}_{i} \sqrt{\eta_i} s_i}}\\
&= \mathbb{E} \PR{\rho_f \eta_i s_i^{*} \mathbf{p}_{i}^H \hat{\mathbf{g}}^{'*}_k \hat{\mathbf{g}}^{'T}_k \mathbf{p}_{i}  s_i}\\
&= \rho_f \text{tr} \PC{\mathbb{E} \PR{ \eta_i \mathbf{p}_{i}^H \hat{\mathbf{g}}^{'*}_k \hat{\mathbf{g}}^{'T}_k \mathbf{p}_{i} s_i s_i^{*}}}\\
&= \rho_f \text{tr} \PC{\eta_{i} \mathbf{p}_{i}^H \hat{\mathbf{g}}^{'*}_k \hat{\mathbf{g}}^{'T}_k \mathbf{p}_{i}}\\
&= \rho_f \eta_i \phi_{k,i}, \\
\end{split} \label{mean_square_2_2}
\end{equation}
and
\begin{equation}
\begin{split}
\mathbb{E} \PR{|A_3|^2} &=  
\mathbb{E} \PR{\left\lvert\sqrt{\rho_f}\tilde{\mathbf{g}}^{'T}_k \mathbf{P}_{\text{MMSE}} \mathbf{N}_{\text{MMSE}} \mathbf{s}\right\rvert^2}\\
&=  \mathbb{E} \PR{\rho_f\mathbf{s}^H \mathbf{N}_{\text{MMSE}} \mathbf{P}_{\text{MMSE}}^H \tilde{\mathbf{g}}^{'*}_k \tilde{\mathbf{g}}^{'T}_k \mathbf{P}_{\text{MMSE}} \mathbf{N}_{\text{MMSE}} \mathbf{s}}\\
&=  \rho_f \text{tr} \PC{\mathbb{E} \PR{\mathbf{N}_{\text{MMSE}} \mathbf{P}_{\text{MMSE}}^H\tilde{\mathbf{g}}^{'*}_k \tilde{\mathbf{g}}^{'T}_k \mathbf{P}_{\text{MMSE}} \mathbf{N}_{\text{MMSE}} \mathbf{s}\mathbf{s}^H}}\\
&= \rho_f \text{tr} \PC{\mathbf{N}_{\text{MMSE}}^2 \mathbf{P}_{\text{MMSE}}^H \mathbb{E} \PR{\tilde{\mathbf{g}}^{'*}_k \tilde{\mathbf{g}}^{'T}_k} \mathbf{P}_{\text{MMSE}}}\\
&= \rho_f \sum_{i=1}^K \eta_i \gamma_{k,i}.
\end{split} \label{mean_square_3_2}
\end{equation}
The final expressions form above include the following definitions
\begin{equation}
    \psi_k = \mathbf{p}_{k}^H \hat{\mathbf{g}}^{'*}_k \hat{\mathbf{g}}^{'T}_k \mathbf{p}_{k}, \text{ for } k= 1,\dots, K,
\end{equation}
\begin{equation}
    \phi_{k,i} = \mathbf{p}_{i}^H \hat{\mathbf{g}}^{'*}_k \hat{\mathbf{g}}^{'T}_k \mathbf{p}_{i},\text{ for } i \neq k, i = 1,\dots,K,
\end{equation}
\begin{equation}
  \boldsymbol{\gamma}_{k} = \text{diag} \chav{\mathbf{P}_{\text{MMSE}}^H\mathbb{E} \PR{\tilde{\mathbf{g}}^{'*}_k \tilde{\mathbf{g}}^{'T}_k} \mathbf{P}_{\text{MMSE}}},
\end{equation}
where $\mathbf{p}_{k} = \PR{p_{1,k},\dots,p_{M,k}}^T$ is the column $k$ of matrix $\mathbf{P}_{\text{MMSE}}$, $\psi_k$ is the $k$th element of vector $\boldsymbol{\psi}$, $\phi_{k,i}$ is the $i$th element of vector $\boldsymbol{\phi}_k$, $\gamma_{k,i}$ is the $i$th element of vector $\boldsymbol{\gamma}_{k}$, and  $\mathbb{E} \PR{\tilde{\mathbf{g}}^{'*}_k \tilde{\mathbf{g}}^{'T}_k}$ is a diagonal matrix with $\PC{(1-n_{m,k})\beta^{'}_{mk}}$ on its $m$th diagonal element.
}
\textcolor{r}{
After substituting \eqref{mean_square_1_2}, \eqref{mean_square_2_2} and \eqref{mean_square_3_2} in the $\text{SINR}_{k,\text{MMSE}}$ expression we obtain
\begin{equation}
\text{SINR}_{k,\text{MMSE}} = \frac{\rho_f \eta_k \psi_k}{\sigma_{w}^2 + \rho_f \sum_{i=1,i \neq k}^{K} \eta_i \phi_{k,i} + \rho_f \sum_{i=1}^K \eta_i \gamma_{k,i}}.
\end{equation}
}

\end{document}